\documentclass[aps,prx,reprint,groupedaddress,superscriptaddress,floatfix,longbibliography]{revtex4-2}

\usepackage{amsmath}
\usepackage{mathtools}
\usepackage{amsfonts}
\usepackage{bbold}

\usepackage{braket}

\usepackage{color}
\usepackage{xcolor}
\usepackage[colorlinks,citecolor=black,linkcolor=blue,urlcolor=blue]{hyperref}

\usepackage{graphicx}

\usepackage{comment}

\def\Egap{\Delta}
\def\Ehop{J}
\def\Clight{c}
\newcommand{\poly}[1]{ \mathrm{poly\,} #1}
\newcommand{\+}{^\dagger}
\newcommand{\phant}{^{\phantom\dagger}}

\def\VA{\mathbf{A}}
\def\VB{\mathbf{B}}
\def\VE{\mathbf{E}}
\def\VJ{\mathbf{J}}
\def\Vsound{v_e}
\def\vel{v}
\def\Vx{\mathbf{x}}
\def\Vy{\mathbf{y}}

\def\destroy{a}
\def\create{\destroy^\dagger}

\def\Nt{N_t}
\def\Ns{N_s}
\def\Nx{N_x}

\def\nint{n_i}
\def\dim{3}
\def\ddim{d}
\newcommand\abs[1]{\left|#1\right|}

\def\Vq{\mathbf{q}}
\def\Vk{\mathbf{k}}
\def\CO{\mathcal{O}}

\begin{document}

\title{Simulating plasma wave propagation on a superconducting quantum chip}

\author{Bhuvanesh Sundar$^\mathparagraph$}
\email[Corresponding author:~]{bsundar@rigetti.com}
\affiliation{Rigetti Computing, 775 Heinz Avenue, Berkeley, California 94710, USA}

\author{Bram Evert$^\mathparagraph$}
\affiliation{Rigetti Computing, 775 Heinz Avenue, Berkeley, California 94710, USA}

\author{Vasily Geyko}
\affiliation{Lawrence Livermore National Laboratory, Livermore, California 94550, USA}

\author{Andrew Patterson}
\affiliation{Rigetti Computing, 138 Holborn, London, EC1N 2SW, United Kingdom}

\author{Ilon Joseph}
\email{joseph5@llnl.gov}
\affiliation{Lawrence Livermore National Laboratory, Livermore, California 94550, USA}

\author{Yuan Shi}
\email{Yuan.Shi@colorado.edu}
\affiliation{Department of Physics, Center for Integrated Plasma Studies, University of Colorado Boulder, Colorado 80309, USA}

\thanks{$^\mathparagraph$These authors contributed equally to this work}

\begin{abstract}
  Quantum computers may one day enable the efficient simulation of strongly coupled plasmas that lie beyond the reach of classical computation in regimes where quantum effects are important and the scale separation is large.
  In this article, we take a first step toward efficient simulation of quantum plasmas by demonstrating linear plasma wave propagation on a superconducting quantum chip. Using high-fidelity and highly expressive device-native gates, combined with an error-mitigation technique, we simulate the scattering of laser pulses from inhomogeneous plasmas. Our approach is made feasible by the identification of a suitable local spin model whose excitations mimic plasma waves, and whose circuit implementation requires a lower gate count than other proposed approaches that would require a future fault-tolerant quantum computer. This work opens avenues to study more complicated phenomena that cannot be simulated efficiently on classical computers, such as nonlinear quantum dynamics when strongly coupled plasmas are driven out of equilibrium.
\end{abstract}
\maketitle

\section{Introduction}
The ability of future fault-tolerant quantum computers to naturally simulate fully entangled quantum dynamics may one day enable simulations of the fundamental forces of nature with unprecedented accuracy and resolution \cite{bauer2023quantum}. 
This will ultimately enable the simulation of high-energy-density plasmas where quantum effects become increasingly important at high energies and/or densities \cite{Drake2018book, grabowski2013wave, graziani2014kinetic, Shi18, Rubin24}. 
In fact, the strongly coupled quantum effects that occur in degenerate fermionic matter also occur in the warm dense matter regime in stars and planetary cores \cite{graziani2014kinetic}, in stopping power and opacity calculations \cite{Rubin24,Shi24jpp}, and in extreme astrophysical plasmas such as those that occur near black holes and neutron stars \cite{uzdensky2014plasma, gueroult2019determining}. 
At a fundamental level, all of these problems are classically hard to simulate (bounder-error quantum polynomial-time complete~\cite{jordan2018bqp}).

The rapid development of quantum computing capabilities has spurred the development of efficient quantum algorithms that target plasma simulation. 
Recent work has developed efficient algorithms for evolving general wave equations \cite{costa2019quantum}, the propagation of waves in plasmas \cite{Engel19, vahala2020unitary, vahala2021one, Dodin21, Novikau2022pra, Joseph23, Novikau2023pra, Novikau2024jpp},
and the underlying physical processes of advection and diffusion \cite{Novikau2025cpc,Novikau2025explicit}.
 Simulating intrinsically quantum plasmas can leverage the significant progress in quantum algorithms for scattering in quantum field theory \cite{jordan2012quantum}, condensed-matter systems \cite{hofstetter2018quantum}, scalar field theory \cite{zemlevskiy2024scalable},
and lattice gauge theories \cite{farrell2025steps}.
In fact, although answering certain questions about the classical dynamics of linear oscillators is intrinsically hard, these problems can be solved efficiently with quantum computers \cite{Babush2023prx}.

While these algorithms generally provide an exponential compression, their implementation on near-term hardware is challenged by the high gate depth required to implement them, typically requiring a gate depth that grows as a polynomial in qubit number. Thus, several studies that have explored the use of noisy intermediate-scale quantum (NISQ) devices for simulating plasma-relevant point examples \cite{Shi21,Zylberman22,Shi24jpp,Porter2025jpp,Novikau2025cpc}
 have shown that fault-tolerant quantum computers will be required for high-precision calculations \cite{Joseph23,Andress2025}.

In this article, we take the first steps toward simulating the physics of wave propagation in quantum plasmas using a strategy that is natural for near-term quantum hardware and apply it to modeling the scattering of electromagnetic waves in plasmas. Typically, solving even a linear quantized scattering problem with a classical computer requires fully solving for the eigenvalues and eigenvectors of the dispersion operator, which requires $\CO(N^\omega)$ operations, where $N$ is the number of discretized spatial points and $2<\omega<3$.
Yet, with a quantum computer we can use Trotterized evolution of a local spin-chain model to efficiently simulate linear plasma wave propagation in a plasma with an inhomogeneous density profile, with no requirement to solve for the eigenvalues or eigenvectors.
As shown in Appendix~\ref{app:advantage}, this leads to a polynomial speedup for simulating the physics of quantum plasmas. Moreover, the model only requires shallow circuits and nearest-neighbor qubit couplings, making it suitable for current devices. 

We present a one-dimensional version of our model, which, as shown in the Appendix~\ref{app:plasma_wave}, can be easily generalized to two, three, or even higher dimensions with a cost that only scales linearly with dimension (the number of interacting neighbors) rather than as a polynomial in the number of qubits. Similar spin-lattice models have already been demonstrated on quantum devices to simulate quantum many-body dynamics \cite{Smith2019simulating, hofstetter2018quantum} and magnetization at infinite temperature \cite{rosenberg2024dynamics};
however, the models in the class studied here have distinct characteristics that make them relevant to plasma physics. While we simulate a solvable spin-chain model in this article, it is possible to add simple extensions to the model with, e.g., $ZZ$ terms in the spin Hamiltonian, which can be used to study classically hard-to-simulate phenomena, such as quantum plasma dynamics including nonlinear and many-body quantum effects. We leave such an experiment to future efforts, and instead focus here on a proof-of-principle demonstration of plasma wave scattering.

We implemented the plasma wave simulations on a nine-qubit sublattice of Rigetti's Ankaa-3 superconducting chip using a strategy that targets universal quantum computation on near-term devices. We compiled our Trotterized evolution operator using the $\sqrt{\textrm{iSWAP}}$-like FSIM 
gates to achieve high-fidelity low gate depth circuits. Unlike previous experiments~\cite{rosenberg2024dynamics} in which targeted gates were calibrated for specific use cases, the native $\sqrt{\textrm{iSWAP}}$-like FSIM gates are highly expressive when combined with single-qubit rotations~\cite{huang2023quantum} and, in general, allow efficient implementation of arbitrary logical circuits. 

While these experiments have relatively forgiving requirements, they still require sophisticated error-characterization and error-mitigation techniques in order to achieve meaningful results \cite{cai2023quantum, temme2017error, van2023probabilistic, filippov2023scalable}. We developed a pseudotwirling technique to twirl the coherent noise in the FSIM gate, and mitigated this noise using Clifford data regression~\cite{czarnik2021error}, which is a heuristic and scalable error-mitigation technique. The overall improvement in fidelity yields the first qualitatively accurate quantum hardware simulation of plasma waves.

\section{Model}\label{sec:model}
We wish to simulate the propagation of electromagnetic (EM) waves in an unmagnetized plasma. These waves, which are coupled oscillations of the electromagnetic fields, including the response of the plasma current, are effectively described by the second-order wave equation
\begin{equation}
\label{eqn:2nd-order wave}
(\partial_t^2-c^2\partial_x^2+\omega_p^2)A=0
\end{equation}
and have the dispersion relation
\begin{equation}\label{eq:plasma dispersion}
\omega_k = \pm \sqrt{\omega_p^2+c^2k^2}.
\end{equation}
Here, $\omega_p$ is the plasma frequency, defined by $\omega_p^2=e^2n/m\epsilon_0$, for particles of charge $e$, mass $m$, and density $n$, and $\Clight$ is the speed of light (or the electron sound speed for electrostatic Langmuir waves). For a magnetized plasma, this dispersion relation is still valid for long wavelength EM modes ({\it L-,R-,O-} and {\it X-}waves \cite{stix1992waves, swanson2020plasma}), but where $\omega_p$ has more complicated dependence on the plasma frequency and the cyclotron frequency, $\omega_c=eB/m$ for magnetic field $B$, for both electrons and ions. In the limit that $\omega_p\rightarrow 0$, the spectrum resembles that of low-frequency long-wavelength ion acoustic waves and magnetohydrodynamic waves in a plasma, where $\Clight$ is replaced by the sound or the Alfv{\'e}n speed.

\begin{figure}[t]
     \centering
         \includegraphics[width=0.7\columnwidth]
         {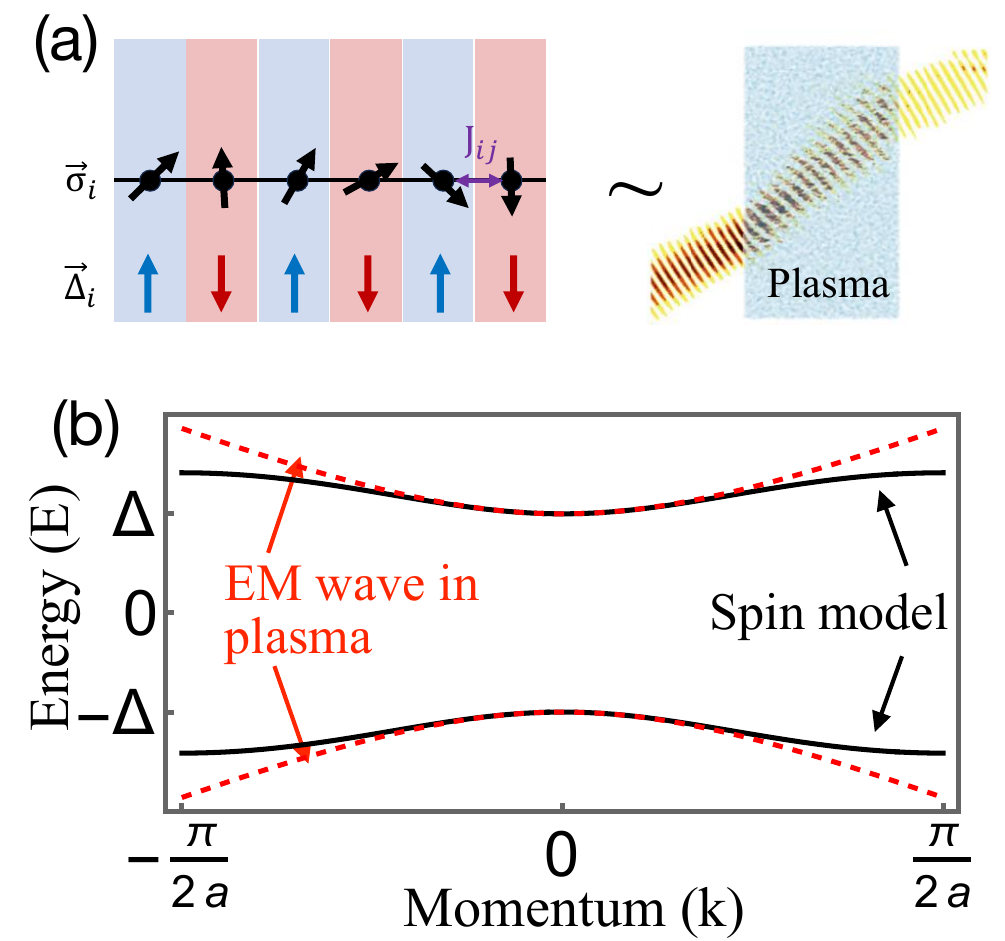}\\
         \caption{We implement a one-dimensional spin model with nearest-neighbor interactions $\Ehop_{ij}$ and a staggered field $\vec{\Egap}_i$ [see Eq.~\eqref{eq:H}], which has the same energy spectrum as that of electromagnetic waves in a plasma, on a superconducting quantum computer. We use this model to simulate linear electromagnetic wave dynamics in a plasma.}
         \label{fig:model}
\end{figure}

We show here that these waves are mimicked by spin-wave excitations in a local spin model given by
\begin{equation}
H = -\frac{\Ehop}{4}\sum_{j=1}^{N-1} (\sigma^x_j \sigma^y_{j+1} - \sigma^y_j \sigma^x_{j+1}) + \frac{1}{2}\sum_{j=1}^N \Egap_j (-1)^j \sigma^z_j.
\label{eq:H}
\end{equation}
where $\sigma^\alpha_i$ are Pauli operators on spin (qubit) $i$. We show in detail in Appendix~\ref{app:plasma_wave} how to derive the spin model that is equivalent to the model of plasma waves. The lattice model of Eq.~\eqref{eq:H} can represent quasiparticles or spins in a condensed-matter system interacting with strength $J$, where the field $\vec \Delta_i$ alternates in direction, as illustrated in Fig.~\ref{fig:model}(a).
The mapping to EM waves in plasmas become apparent when we write the equation of motion for singly excited states $\ket{0\cdots1_j\cdots0}$. Writing the amplitude of the local excitation as $b_j$, the equation of motion for these amplitudes is
\begin{equation}
    \label{eq:ODE}
    i\hbar\partial_t b_j = \frac{i}{2}\Ehop(b_{j-1}-b_{j+1}) +\Egap_j(-1)^j b_j.
\end{equation}
Differentiating again and assuming uniform $\Egap_j$ gives 
\begin{equation}
\hbar^2\partial_t^2 b_j=\frac{1}{4}\Ehop^2(b_{j+2}-2b_j+b_{j-2}) - \Egap^2 b_j,
\end{equation}
which is a central-difference discretization of Eq.~\eqref{eqn:2nd-order wave}. The degrees of freedom at the odd and even sites of the lattice model can be interpreted as linear combinations of the electric field and current in a domain with half the total number of grid points. For uniform $\Egap_j = \Egap$, this discretized wave equation can be solved exactly: the eigenmodes have the dispersion relation~\footnote{The standard way to solve this Hamiltonian is by mapping it to a free-fermionic Hamiltonian or a hardcore bosonic Hamiltonian; See Appendix~\ref{app:exact_soln} for details.} 
\begin{equation}\label{eq:lattice dispersion}
\hbar\omega_k = \pm \sqrt{\Egap^2 + \Ehop^2\sin^2 ka}.
\end{equation}
where $a$ is the lattice spacing. At long wavelength, $\hbar\omega_k \simeq \pm \sqrt{\Egap^2+(\Ehop ka)^2}$ is the same dispersion relation as Eq.~\eqref{eq:plasma dispersion}, and is illustrated in Fig.~\ref{fig:model}(b).

Thus, we can efficiently simulate the linear dynamics of plasma waves on quantum hardware using Trotterized evolution of $H$. By varying $\Egap_j$, e.g. making it uniform, having sharp jumps, or having a spatial profile, we can achieve full-wave simulations of linear wave dynamics in uniform plasmas, ducted plasmas, or inhomogeneous plasmas. In this paper, we consider simulation of all three scenarios.

\begin{table}[t]
    \centering
    \begin{tabular}{||r | c||}
        \hline
        Metric & Value \\
        \hline\hline
        T1 & $ 37.6 \mu s$\\
        \hline
        T2 & $ 22.3 \mu s$ \\
        \hline
        FSIM Infidelity & $0.60\% $\\
        \hline
        FSIM Duration & $54 ns$ \\
        \hline
        RX($\pi$/2) Infidelity & $ 0.09\% $ \\
        \hline
        RX($\pi/2)$ Duration & $ 20 ns $ \\
        \hline
    \end{tabular}
    \caption{Summary metrics of the sublattice of the Ankaa-3 QPU. The $T_1$ and $T_2$ values are measured individually, and the gate fidelities are mean values reported in the circuit context.}
    \label{tab:ankaa-3-char}
\end{table}

\section{Experiment} We perform our experiments on a nine-qubit sublattice of Rigetti's Ankaa-3 chip which is composed of transmon qubits arranged on a square grid. The qubits are connected by tunable couplers, which, when activated, realize an interaction between the qubits~\cite{sete_floating_2021}. The interaction, which is predominantly a transverse interaction with an additional small longitudinal interaction in the two-qubit Hilbert space, realizes the so-called FSIM gate~\cite{rosenberg2024dynamics} up to one-qubit phases (see Appendix~\ref{app:hardware}). We calibrate the duration of the interaction to target a gate that is close to a $\sqrt{\textrm{iSWAP}}$ up to one-qubit phases. The $\sqrt{\textrm{iSWAP}}$ is a highly expressive gate that can be used to efficiently compile arbitrary two-qubit gates~\cite{huang2023quantum}. Since the $\sqrt{\textrm{iSWAP}}$ gate has a shorter duration than the iSWAP gate, it has reduced decoherence and thus a higher fidelity than iSWAP. On each edge of our quantum processor, we calibrate and learn the gate being realized using process tomography and a variant of cross-entropy benchmarking, as detailed in Appendix~\ref{subsec:unitary_learning}. We implement arbitrary single-qubit unitary operations using four phased microwave pulses known as the ``PMW4'' decomposition \cite{chen2023compiling}. The error per layered gate (EPLG) \cite{mckay2023benchmarking} in the relevant nine-qubit sublattice on Ankaa-3 was found to be 1.48\% (see Table~\ref{tab:ankaa-3-char} for other performance data). Using these gates, we express the target logical circuits using an approximate numerical compilation technique, as explained in Appendix~\ref{subsec:numerical_compilation}. As we describe below, the logical circuits we target realize Trotterized evolution with Eq.~\eqref{eq:H}. The resulting experimental circuit is illustrated in Fig.~\ref{fig:sm-trotter}.

We mitigate hardware noise in the experiment using a combination of twirling and a linear regression technique, similar to Ref.~\cite{czarnik2021error}. The twirling method that is commonly used in quantum circuits with Clifford two-qubit gates is Pauli twirling, and it transforms any Markovian noise that is closed within the qubit's computational space to stochastic Pauli noise, which is amenable to error mitigation; however, the FSIM gate that we used in our experiment is not a Clifford gate, and therefore it cannot be twirled using Pauli gates. We therefore developed a pseudotwirling technique for the FSIM gates that converts Markovian coherent noise within the computational space, other than errors in the calibration of the FSIM parameters, to stochastic Pauli noise (see Appendix~\ref{subsec:twirling_fsim} for details). We then apply the linear regression technique to mitigate the effects of the now mostly stochastic noise (see Appendix~\ref{subsec:cdr}). Additionally, where applicable, we rescale the observables such that the expectation value of $\sigma^z_{\rm tot}$ is conserved (see Appendix~\ref{subsec:Zconservation}).

\begin{figure}[t]
     \centering
         \includegraphics[width=1.05\columnwidth]{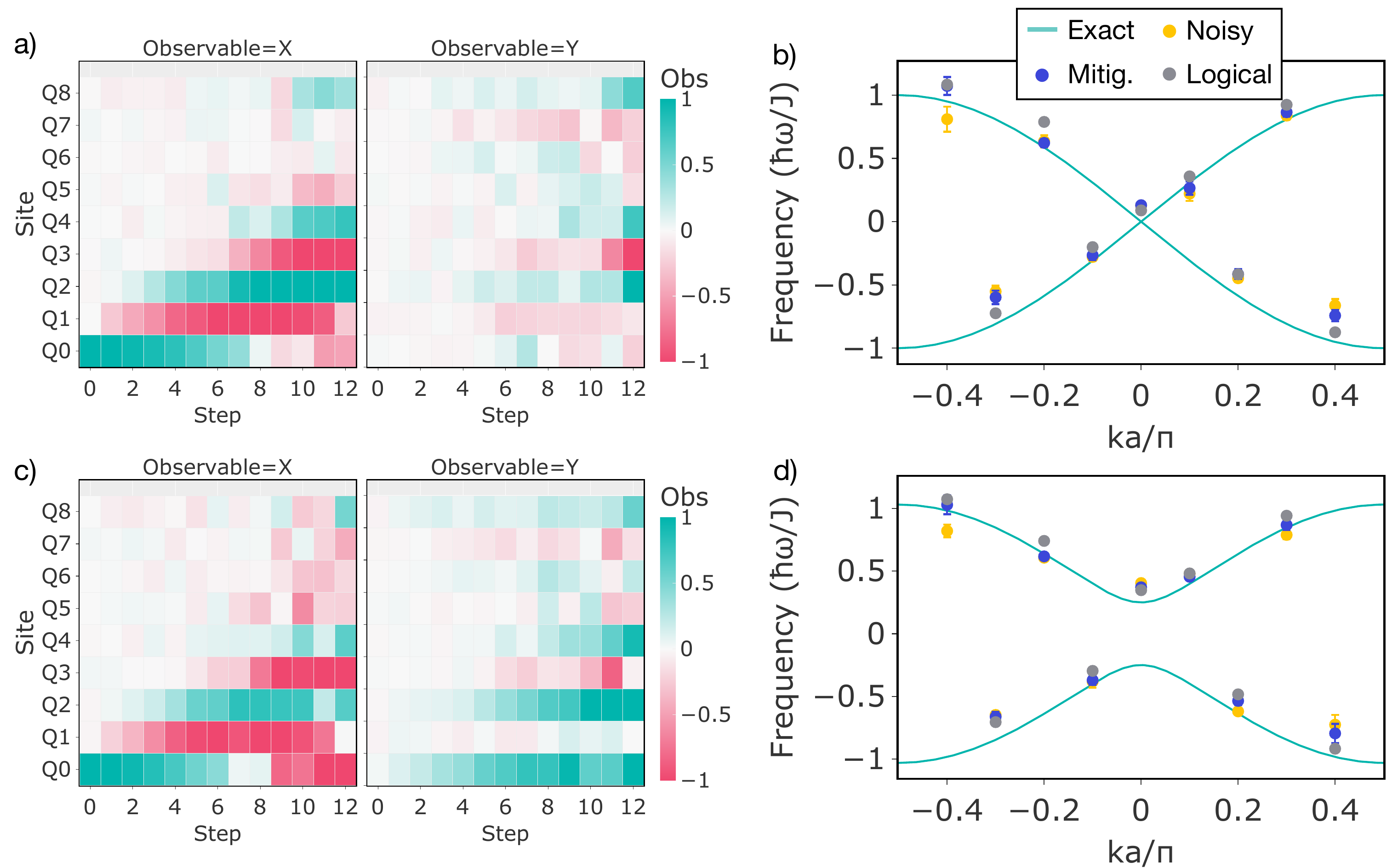}
         \caption{The dispersion relation of the spin Hamiltonian [Eq.~\eqref{eq:H}] is analogous to waves in plasmas. The spectrum is measured via a many-site Ramsey-type experiment, where the time evolution of the complex phase of local spin observables reveals the energies of the eigenmodes. We consider (a-b) a low density plasma ($\Egap\rightarrow0$) and (c-d) $\Egap=\Ehop/4$. Panels (a,c) show the error-mitigated observables versus time; panels (b,d) show the extracted excitation spectrum from noisy (yellow) and mitigated (blue) data, along with the exact spectrum (teal) and the expected spectrum from a noiseless simulation of the circuit (gray).}
         \label{fig:dispersion-expt}
\end{figure}

At the end of all of the circuits, we obtain classical shadows of the state via measurements of all the qubits in random bases~\cite{huang2020predicting}. Classical shadows is a powerful technique for estimating several observables to high accuracy using relatively few shots, and, importantly here, it also effectively acts as readout-error symmetrization due to the randomization of the measurement basis. The random basis for measurement on individual shots is triggered by a pseudorandom number generator on the quantum chip's control system, which enables $10^6$ shots in a few seconds.

\begin{figure}[t]
\includegraphics[width=1.0\columnwidth]{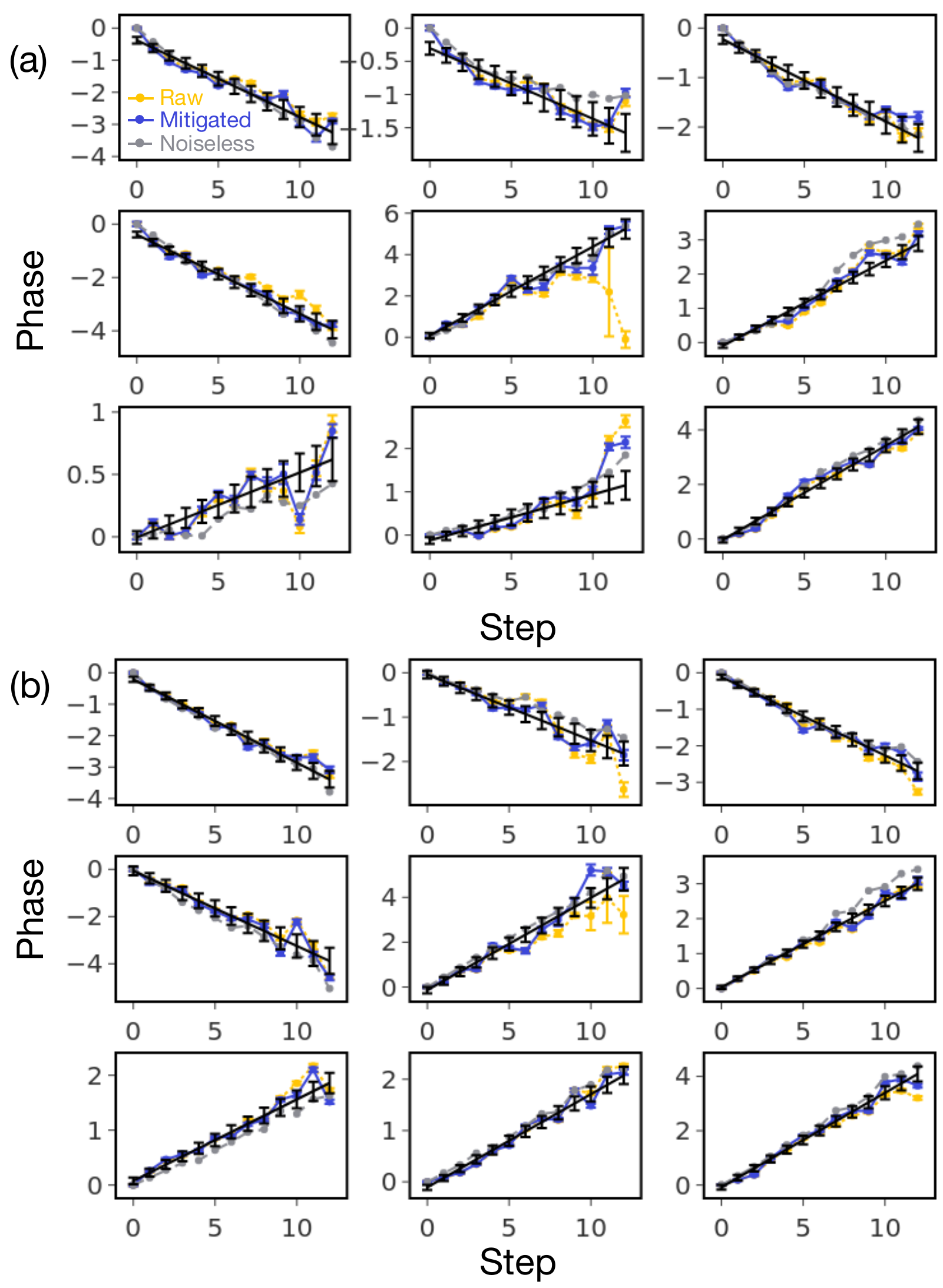}
\caption{Phase evolution of the eigenmodes of the system versus time for (a) $\Egap=0$, and (b) $\Egap=\Ehop/4$. Magenta, teal, and yellow points plot the complex phase of the eigenmodes obtained from error-mitigated data, the raw experimental data, and a noiseless simulation of the circuit. Black lines show linear fits to the error-mitigated data. Other colored lines are guides to the eye. Each panel corresponds to one eigenmode $q$. The eigenfrequency of each eigenmode is obtained by applying a linear fit to the error-mitigated data.}
\label{fig:phase_evolution}
\end{figure}

\begin{figure*}[thb]
     \centering
     \includegraphics[width=1.05\textwidth]{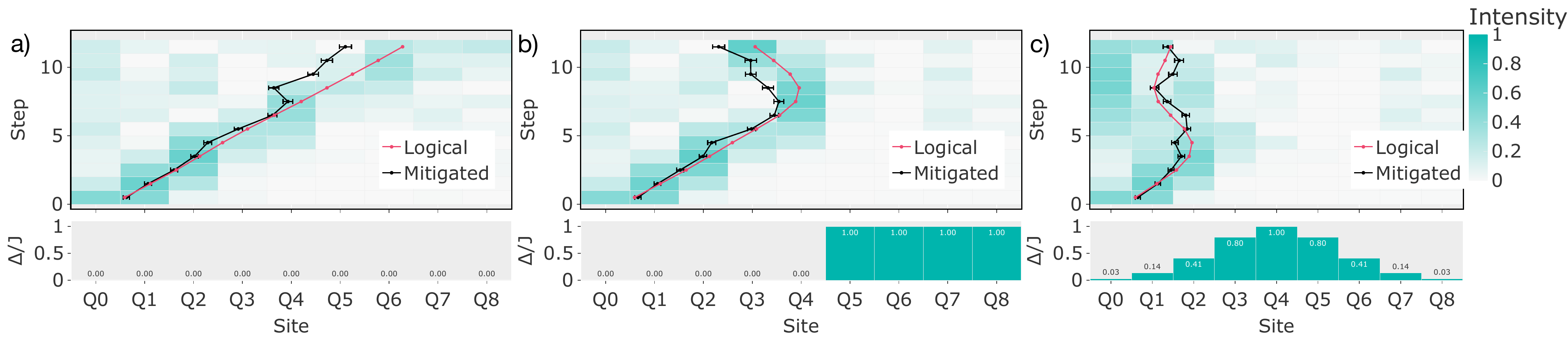}
     \caption{An electromagnetic wave packet propagating (a) in vacuum ($\Egap = 0$), (b) from vacuum to a sharp jump in plasma density (which mimics the edge of a confined overdense plasma), and (c) through an inhomogeneous overdense plasma with a Gaussian density profile. In each case, the profiles of plasma frequency are shown on the bottom, and the intensity of the propagating wave packet is shown on the top. In (a), the wave packet propagates nearly ballistically until it approaches the edge, where the boundary condition is reflective. In (b) and (c), the wave packet propagates until it approaches the sharp jump or inhomogeneous profile and mostly reflects back. The black lines show the center of mass (CoM) of the wave packet obtained from the mitigated experimental data, and the red lines show the wave packet's CoM from a noiseless simulation of the experiment.
     }
    \label{fig3}
\end{figure*}

\section{Measuring the dispersion relation}
We design our first experiment to measure the energies of the spin-wave excitations, using a many-body Ramsey-type scheme.

We initialize one qubit in the superposition $(\ket{0}+\ket{1})/\sqrt{2}$ and 
all remaining qubits in the $\ket{0}$ state. This initial state is a superposition of the vacuum of spin waves, and of spin waves at all wave vectors $k$ with appropriate coefficients. 
Denoting $\ket{\rm vac}$ as the state with all qubits as $\ket{0}$, which is the vacuum of spin waves, $\ket{i} \equiv \sigma^+_i\ket{\rm vac}$ as the state with qubit $i$ in $\ket{1}$ and all others as $\ket{0}$, and  $\ket{k}$ as the spin wave with wave vector $k$, the initial state is
\begin{equation}
\ket{\psi(t=0)} = \frac{\ket{\rm vac} + \ket{i}}{\sqrt{2}} = \frac{\ket{\rm vac} + \sum_k c_{ik}\ket{k}}{\sqrt{2}}.
\end{equation}
Here, $c_{ik}$ are the real-space amplitudes of the spin wave with wave vector $k$ (see Appendix~\ref{app:exact_soln} for the exact solutions of the spin waves).

We then realize first-order Trotterized evolution with $H$ on the chip, using a Trotter time-step size $\delta=0.8$, wherein each wave vector component $k$ accumulates phase at the rate $\omega_k$. The vacuum state remains as a reference, having no phase evolution.
After time evolution, the wave function is
\begin{equation}
\ket{\psi(t)} = \frac{\ket{\rm vac} + \sum_k c_{ik}\phant \ket{k} e^{-i\omega_kt}}{\sqrt{2}}.
\end{equation}
We measure $\braket{\sigma^x_j} + i\braket{\sigma^y_j}$, which is sensitive to the relative phases accumulated between the vacuum and the spin waves during Trotterized evolution.
The expectation value of $\sigma^x_j + i\sigma^y_j$ is
\begin{align}
\braket{\psi(t) \vert \sigma^x_j + i\sigma^y_j \vert \psi(t)} =& \frac{1}{2}\sum_k c_{ik}e^{i\omega_k t} \braket{k \vert j} \nonumber\\
=& \frac{1}{2}\sum_k c_{ik}\phant c_{jk}^* e^{i\omega_k t}.
\end{align}
We perform the experiment with ($\Egap>0$) and without ($\Egap=0$) a plasma. We remark that $(\sigma^x_j + i \sigma^y_j)$ is not a Hermitian observable, but we can measure its real part $\sigma^x_j$ and imaginary part $\sigma^y_j$ separately.

Figures~\ref{fig:dispersion-expt}(a) and~(c) show the localized spin-wave packet spreading through the lattice in the cases $\Egap=0$ and $\Egap=\Ehop/4$, respectively. 
We then extract the frequency $\hbar\omega_q$ from the complex phase evolution of $\sum_j \braket{\sigma^x_j + i\sigma^y_j} c_{jq}$. We know the amplitudes $c_{jq}$ for the eigenmodes of this Hamiltonian (see Appendix~\ref{app:exact_soln}); therefore, we are able to extract the spectrum of $H$ in this way. Figure~\ref{fig:phase_evolution} shows that the complex phase of the eigenmodes evolves approximately linearly with time. The rate of phase evolution is the eigenfrequency of the modes.

The resulting spin-wave excitation spectra are shown in Figs.~\ref{fig:dispersion-expt}(b) and (d) for $\Egap=0$ and $\Egap=\Ehop/4$, respectively. The teal lines show the exact frequencies of the spin-wave excitations. The gray dots are obtained by extracting the frequency spectrum from a noiseless simulation of the circuit, using the same postprocessing procedure. There is a small difference between the exact values (teal curve) and the noiseless simulation values (gray dots) due to Trotter error in the evolution. Finally, the blue and yellow dots are the frequency spectra from experimental data with and without error mitigation, respectively. The frequency spectrum obtained from experimental data agrees excellently with that from the noiseless simulation and with the exact eigenfrequencies.

In particular, the experimental data show the presence of a mode at $\omega=0$ when $\Egap=0$ [Fig.~\ref{fig:dispersion-expt}(b)], whereas there is no mode at $\omega=0$ when $\Egap>0$ [Fig.~\ref{fig:dispersion-expt}(d)]. In fact, the spectrum for $\Egap>0$ has an energy gap. The energy gap is the plasma frequency $\omega_p$, which is determined by the plasma density; only EM waves with frequency greater than $\omega_p$ propagate in the plasma.

In the absence of wave-wave interactions, this method represents a highly sensitive scalable test of the ability to accurately reconstruct each eigenmode. Extracting the spectrum of a general nonsolvable Hamiltonian may require sophisticated techniques like quantum phase estimation.

\section{Propagation of waves in the plasma}\label{sec:plasma_wave_propagation}
Modeling the propagation of EM waves in plasmas with complex geometries is a computationally expensive, but important, task in many plasma-relevant scenarios. Here, we take the first step toward this goal on a quantum computer, by simulating propagation of EM waves in plasma with increasingly more general density profiles. In particular, we consider propagation in three scenarios -- vacuum ($\Egap= 0$), a plasma with a sharp jump in density, and an inhomogeneous plasma with a smooth density profile. While the first two scenarios are well understood analytically, they serve as control experiments to benchmark and validate our simulation. The third scenario is a proof-of-principle demonstration of how a quantum computer could be used to simulate real EM wave propagation in more realistic plasmas. 

In our experiments, the spin-wave packet, $\ket{\psi} = \sum_i \alpha_i \ket{i}$, is prepared with an entangling operation compiled into our native gates. This wave packet has exactly one spin-wave excitation, distributed over multiple sites in real space, and multiple $k$ in wave-vector space. The mean $k$ is dictated by the complex phases of $\alpha_i$, and the spread over $k$ is dictated by the inverse of the spread in real space. In a large-scale fault-tolerant experiment, a broad real-space wave packet may be prepared such that it is concentrated in $k$. Here, we initialize a two-site wave packet. The relative complex phase between $\alpha_0$ and $\alpha_1$ is 0; therefore, the initial condition is a superposition of left- and right-propagating waves. We launch the wave near the left boundary from the region with $\Egap=0$. Because we use a reflective boundary condition, the net effect is that a wave packet is launched with a group velocity $v = \Ehop/a\hbar$ that is positive.

Figure~\ref{fig3} shows the wave propagation in the three scenarios. In the first case, Fig.~\ref{fig3}(a), the plasma density is zero; therefore, the wave is the vacuum EM wave. The solid curves plot the center of mass of the wave packet. In a noiseless simulation (red), the wave packet propagates freely at its group velocity. Deviation from a constant speed occurs at later times due to finite size of the lattice. The experimental data track this propagation qualitatively well until at least half the lattice.

In Figs.~\ref{fig3}(b) and (c), the wave packet encounters a sharp jump in the plasma density and an inhomogeneous plasma density profile, respectively. The maximum plasma barrier, $\Egap=\Ehop$, is larger than the wave packet's vacuum frequency, and it therefore corresponds to an overdense plasma. In both scenarios, most of the wave packet is reflected because the plasma is overdense. The error-mitigated experimental data show good qualitative agreement with the noiseless numerical simulation, as shown by the qualitative agreement in the traces of their centers of mass.

\begin{figure}[th]
\centering
\includegraphics[width=1.0\columnwidth]{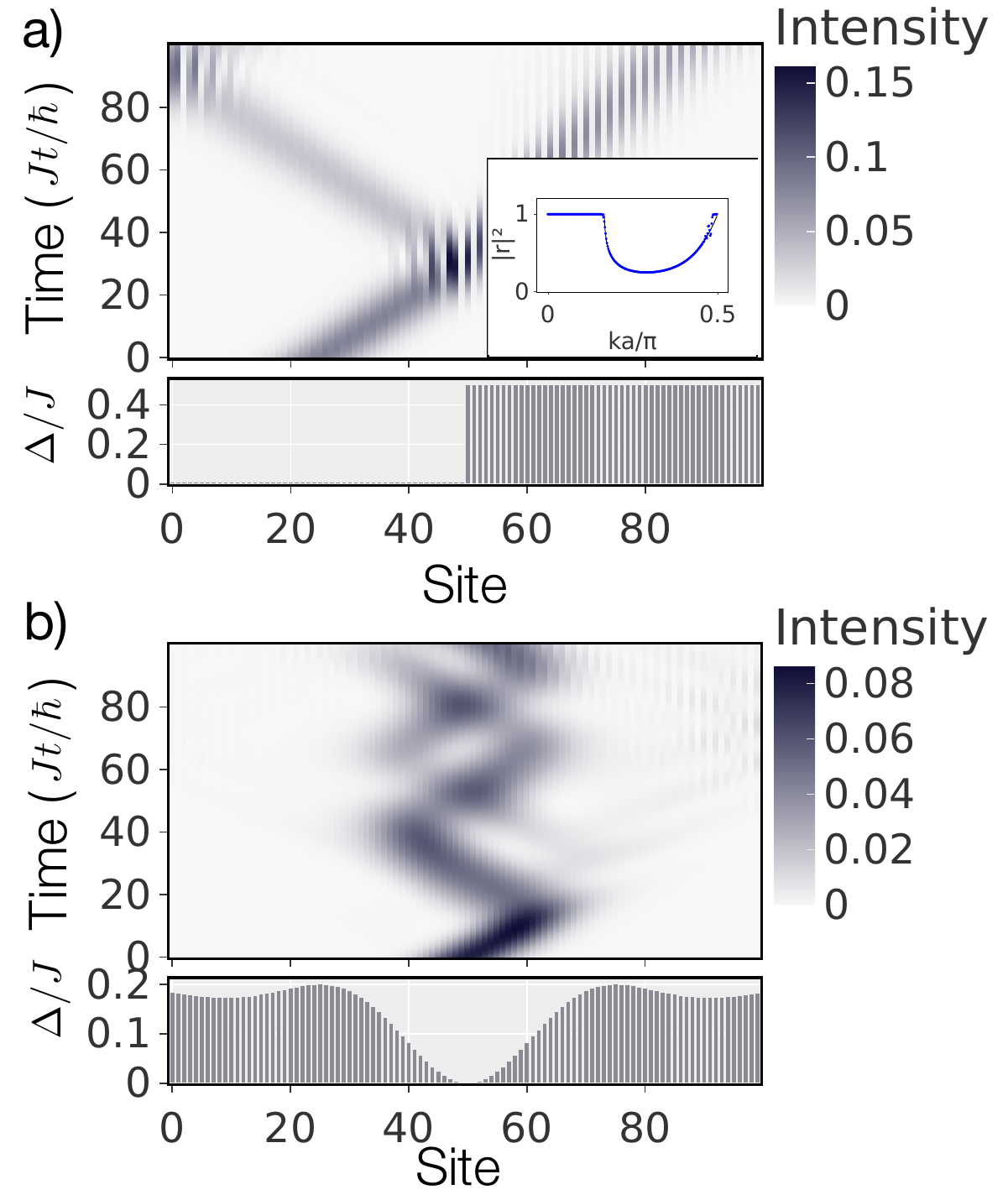}
\caption{Classical simulations of EM waves propagating in various mass profiles: (a) reflection from a sharp boundary and (b) propagation through a more complicated profile. The inset in (a) shows the intensity of the reflected wave, $|r|^2$, versus $k$. The solid line shows the analytically predicted reflection (see Appendix~\ref{app:reflection_coefficient}) and the points show the numerically computed values.}
\label{fig:large}
\end{figure}

State-of-the-art full-wave simulations use $\CO10^6$-$10^8)$ spatial grid points on classical computers. While current qubit error rates limit us to simulations of plasma wave propagation on nine qubits, which corresponds to a spatial discretization on nine grid points, smaller error rates that allow experiments on larger lattices are within reach in the near term. Fault-tolerant computers, which would be needed for large-scale quantum simulations, are also rapidly advancing. With this motivation, in Fig.~\ref{fig:large}, we present a classical emulation of the lattice model for a large system with 100 sites. The Hilbert-space size for simulating a single wave packet propagating in a plasma scales in proportion to the number of grid points $n$ and is therefore feasible to simulate classically.

\section{Quantum advantage}\label{sec:advantage}
Quantum computers are naturally able to simulate quantum dynamics and, thus, can offer significant quantum advantage. Here, we show a brief calculation of the expected quantum advantage for simulating propagating plasma waves, with more details given in Appendix~\ref{app:advantage}.

The model that approximates the quantum field theory of linear electromagnetic wave propagation in the low photon (or plasmon) density limit is the noninteracting bosonic 
model in Eq.~\eqref{eqn:noninteracting bosons}, which is an integrable model. For a lattice with $N_s$ sites, the evolution of the noninteracting model can be classically computed in time
$C = \CO(M_0 K N_s^2) \text{--} \CO(M_0 N_s^3)$, where $M_0$ is a constant that depends on the initial conditions and/or the number of single-particle observables and $K$ represents the sparsity of the matrix or the number of Krylov iterations.
In fact, in one dimension (1D), it only takes $C = \CO(M_0 K N_s^2)$.
The dominant contribution to the classical cost is 
either exact diagonalization or exponentiation of the single-particle Hamiltonian. Therefore, classical simulation is feasible if the initial condition can be expressed in terms of a polynomial number of single-particle states or if a polynomial number of single-particle observables is desired.

Since the quantum computer does not need to explicitly diagonalize the Hamiltonian, it can still provide significant advantage in these cases. The quantum algorithm's cost only scales with the number of time steps, $\Nt$, which is proportional to $\Nx$ due to the Courant-Friedrichs-Lewy criterion for stability and accuracy, $Q \propto \CO(d^2 N_x)$, where $d$ is the number of spatial dimensions. Therefore, the quantum speedup range is 
\begin{align}
C/Q=\CO(M_0 K \Nx^{2\ddim-1}/\ddim^2) 
\, \text{to}\,
\CO(M_0 \Nx^{3\ddim-1}/\ddim^2).
\end{align}
Nonetheless, a practical advantage may not be realizable on currently feasible system sizes;
 we discuss cases where quantum advantage may still be achievable in Appendix~\ref{app:advantage}.

A more general Hamiltonian with nonlinear interactions, e.g. with additional ZZ terms, does not map to a noninteracting model and, hence, cannot be classically simulated in $\poly(\Ns)$ time. Solving this system classically generically takes exponential time in $\Ns$, scaling as $\CO(2^{3\Ns})$ in the worst case. 
Thus, the case of nonlinear interactions is also promising for realizing quantum advantage.

\section{Summary} 
We simulate the scattering of plasma waves from an inhomogeneous medium using a quantum device. Our spin-lattice model represents quantum plasma waves using $n$ qubits for $n$ spatial grid points, which results in a circuit with a shallow gate depth of $\CO(n)$ that can be run on NISQ devices. While we have only used singly excited spin states, which form an $n$-dimensional subspace of the $2^n$-dimensional Hilbert space, to represent linear plasma waves, more complicated spin excitations can be simulated just as efficiently on quantum computers. With more general spin Hamiltonians, for example, with additional $ZZ$ interactions, our approach can be used to simulate nonlinear effects in plasmas, such as electromagnetically induced transparency~\cite{harris1996electromagnetically}, laser-plasma scattering, and modulational instabilities~\cite{Shi24jpp}.

\begin{acknowledgments}
This material is based upon work supported by the U.S. Department of Energy, Office of Science, under Award no. DE-SC0021661. This publication was prepared to include an account of work sponsored by an agency of the United States Government. Neither the United States Government nor any agency thereof, nor any of their employees, makes any warranty, express or implied, or assumes any legal liability or responsibility for the accuracy, completeness, or usefulness of any information, apparatus, product, or process disclosed, or represents that its use would not infringe privately owned rights. Reference herein to any specific commercial product, process, or service by trade name, trademark, manufacturer, or otherwise does not necessarily constitute or imply its endorsement, recommendation, or favoring by the United States Government or any agency thereof. The views and opinions of authors expressed herein do not necessarily state or reflect those of the United States Government or any agency thereof. 
The work by Lawrence Livermore National Laboratory was performed under the auspices of the U.S. Department of Energy (DOE) under Contract No. DE-AC52-07NA27344. 
I.J. and V.G. were supported by the DOE Office of Fusion Energy Sciences projects SCW1736 and SCW1680. Y.S. is supported in part by U.S. Department of Energy under Grant No. DE-SC0020393. B.S. and B.E. thank Mark J. Hodson, Maxime Dupont, and Tyler Wilson for valuable discussions.

I.J. and B.S. developed the spin chain model to emulate waves in plasmas and derived the protocol for measuring the dispersion relation. Y.S. conceived of the scattering experiments performed in this work. B.E. ran the experiments on Rigetti's quantum chips. A.P. and B.E. developed the error-mitigation techniques used in the experiment. B.S. provided theory support, analyzed the experimental results, and wrote the manuscript with contributions and editing from all authors.
\end{acknowledgments}

B.S., A.P., and B.E. are, have been, or may in the future be participants in incentive stock plans at Rigetti Computing Inc.

\section*{Data availability}
All the relevant data created from experiments in this work are publicly available at \hyperlink{doi.org/10/5281/zenodo.16115660}{doi.org/10/5281/zenodo.16115660} \cite{data_availability}.

\appendix

\section{Mapping Plasma Waves to Spin Chains}\label{app:plasma_wave}
Here, we show how to derive the spin model [Eq.~\eqref{eq:H}] to simulate a continuum plasma model.

\subsection{Continuum model}

Plasma dynamics is determined by Maxwell's equations,
\begin{align}
\partial_t \VB&=-\nabla\times \VE & \nabla\cdot\VB&=0
\\
\partial_t \VE&=\Clight^2 \nabla\times \VB -\frac{\VJ}{\epsilon_0} &\nabla\cdot\VE&=\frac{\rho}{\epsilon_0}
\end{align}
and the evolution of the charge carriers in response to the electromagnetic fields, where $\epsilon_0$ is the electric permittivity.
For an unmagnetized plasma composed of light electrons and heavier ion species, the linearized electric current evolves via
\begin{align}
\partial_t \VJ= -\nabla\Vsound^2\rho+ \epsilon_0\omega_p^2\VE
\end{align}
where $\omega_p$ is the plasma frequency and $\Vsound^2=\gamma T_e/m_e$ sets the electron sound speed.
Combining these equations leads to the wave equation for a massive vector field
\begin{align}
\partial^2_t \VE+ \Clight^2 \nabla\times \nabla\times \VE =-\frac{\partial_t\VJ}{\epsilon_0}=-\omega_p^2\VE + \nabla\Vsound^2\nabla\cdot\VE.
\end{align}
Assuming that the coefficients are constant, then in radiation gauge, $\VE=-\partial_t\VA$, integrating in time gives
\begin{align}\label{eqn:wave_eqn}
\partial^2_t \VA+ \Clight^2 \nabla\times \nabla\times\VA +\omega_p^2\VA - \nabla\Vsound^2\nabla\cdot\VA=0.
\end{align}
These equations can be derived from the Hamiltonian
\begin{align} \label{eq:H_plasma-wave}
H = \frac{ \epsilon_0}{2}\int d^{\dim}x\left(\abs{\VE}^2 + \abs{\Clight \nabla\times\VA}^2 + \abs{\Vsound \nabla\cdot\VA}^2+\abs{\omega_p \VA}^2\right).
\end{align}
There are three modes: two electromagnetic waves with polarization perpendicular to the wave vector that can travel near the speed of light and one electrostatic Langmuir wave with longitudinal polarization along the wave vector that can travel near the electron sound speed.

In the main text, we assume that the spatial variation in the plasma is along the direction of wave propagation.
This causes the polarization to decouple from the evolution and, hence, to remain fixed in time.
Thus, we can treat the polarization of each mode individually.
In this case, Eq.~\eqref{eqn:wave_eqn} can be written as three one-dimensional Klein-Gordon equations, one for each vector component,
\begin{equation}\label{eqn:KG}
\partial^2_t A_\mu +\omega_p^2 A_\mu - \vel^2 \partial_x^2 A_\mu = 0,
\end{equation}
where we assume that the direction of wave propagation is $\hat x$. Here, $\vel$ is the velocity of the wave with polarization along $\mu$, in which $\vel=\Clight$ is the speed of light for electromagnetic waves when $\mu$ is perpendicular to $x$, and $\vel=\Vsound$ is the electron sound speed for Langmuir waves when $\mu = x$.

Canonical quantization for bosonic fields leads to the equal-time canonical commutation relations (CCR) $\epsilon_0[E_\mu^\dagger(\Vx),A_\nu(\Vy)]=i\hbar \delta_{\mu\nu}\delta^3(\Vx-\Vy).$
In general, the solutions decompose into linear eigenfunctions labeled by the discrete indices, $\mu$ for the polarization of the different modes, and $\Vk$, representing spatial degrees of freedom. 
In a uniform plasma, $\Vk$ is the wave vector, but in a nonuniform plasma, this is simply an index over all eigenstates for each mode, which is discrete for a bounded spatial domain.
We define creation and destruction operators that obey the CCR $\left[\destroy_{\mu,\Vk}, \create_{\nu,\Vq}\right]=\delta_{\mu\nu}\delta_{\Vk,\Vq}$
via
\begin{align} \label{eq:define-create-destroy}
\create_{\mu,\Vk}&=\frac{\alpha_{\mu,\Vk}}{\sqrt{2\abs{\hbar\omega_\Vk/\epsilon_0}}}
&
\destroy_{\mu,\Vk}=\frac{\alpha^*_{\mu,\Vk}}{\sqrt{2\abs{\hbar\omega_\Vk/\epsilon_0}}}
\end{align}
where $\alpha_{\mu,\Vk}$ represents the projection of each orthonormal eigenfunction $\boldsymbol{\phi}^*_{i,\Vk}(\Vx)$ onto the fields
\begin{align} \label{eq:eigenmodes}
\alpha_{\mu,\Vk}=\int d^\dim \Vx\ \boldsymbol{\phi}^*_{\mu,\Vk}(\Vx)\cdot \left[\VE^*(\Vx)-i\abs{\omega_\Vk}\VA(\Vx)\right].
\end{align}
This leads to the Hamiltonian
\begin{align}
H = \sum_{\mu,\Vk} \hbar\abs{\omega_{\mu,\Vk}} \create_{\mu,\Vk} \destroy_{\mu,\Vk} + H_0.
\end{align}
where $H_0$ is the vacuum energy of the plasma model.

In what follows, we will write a bosonic lattice model that produces the dynamics of Eq.~\eqref{eqn:KG} and map that bosonic model to a spin model; however, this is made nontrivial by the fact that while $\destroy_{\mu,\Vk}$ satisfies the standard bosonic commutation relation, the vector potential $\VA$ does not. Nevertheless, one can show that $\destroy_\mu(x) \equiv \sqrt{\epsilon_0/(2\hbar)} \left((f \ast E_\mu^*)(x) + i(g \ast A_\mu)(x)\right)$ also satisfies Eq.~\eqref{eqn:KG}, where $\ast$ is the convolution operator, and $f$ and $g$ are Fourier transforms of $1/\sqrt{|\omega_k|}$ and $\sqrt{|\omega_k|}$, respectively. Therefore, we will derive our bosonic lattice model to simulate Eq.~\eqref{eqn:KG} for $\destroy_\mu(x)$.

\subsection{Mapping to local spin model}
Due to the limited resources of present-day hardware platforms, we would like to simulate the plasma wave equation using a local spin model that is naturally represented by the qubits of a quantum computer. 
Hence, we limit the occupation number at each point in space to two possibilities, 0 or 1.
The qubit Hamiltonian is first order in momentum, so it cannot directly represent the plasma Hamiltonian, which is second order in momentum, and still maintain the correct bosonic quantization conditions.
Moreover, generating the second-order dispersion relation for plasma waves requires a kinetic term that must be approximated as a first-order differential operator.
For real bosons, this requires at least two qubits per lattice site, e.g. a complex bosonic field.

One approach is to collect the components of the field, represented by qubits or spins, in the form of a Dirac spinor, while retaining bosonic statistics. 
For each mode with velocity $v$, we can write the Dirac equation for the Dirac spinor $\Psi(t,\Vx)$ as
\begin{align} \label{eqn:Dirac}
i\gamma^0 \partial_t \Psi = -i\vel \gamma^1\partial_x \Psi + \omega_p\Psi,
\end{align}
where $\gamma^j$ are Dirac matrices satisfying $(\gamma^0)^2 = -(\gamma^1)^2 = 1$ and $\{\gamma^0, \gamma^1\} = 0$, and $\Psi(x)$ is a spinor. One is free to use any representation of $\gamma^\mu$ provided that it satisfies these anticommutation relations. Multiplying both sides of Eq.~\eqref{eqn:Dirac} by $i\gamma^\mu \partial_\mu$ gives the Klein-Gordon equation in Eq.~\eqref{eqn:KG} for $\Psi$. Since both $\Psi(x)$ and $\destroy_\mu(x)$ obey Eq.~\eqref{eqn:KG}, a natural conclusion is that $\Psi(x) \propto \destroy_\mu(x)$.

The spatially discretized form of the Dirac equation is
\begin{align} \label{eqn:Dirac_discretized}
i\gamma^0 \partial_t \Psi(x) = -i\frac{\vel}{2a}\gamma^1\left[ \Psi(x+a)-\Psi(x-a) \right] + \omega_p\Psi(x).
\end{align}
Without loss of generality, we set $\gamma^0 = -\tau^z$ and $\gamma^1 = i \tau^y$. Then, the spinor $\Psi(x)$ must be defined in terms of $\destroy_\mu(x)$ such that $\partial^2_t \destroy_\mu$ is given by Eq.~\eqref{eqn:KG}. Choosing $\Psi(2x) = \destroy_\mu(2x) (1 \ 0)^T$ on even sites and $\Psi(2x+a) = \destroy_\mu(2x+a) (0 \ 1)^T$ on odd sites accomplishes this. The two components of the spinor equation [Eq.~\eqref{eqn:Dirac_discretized}] can be written as two separate equations,
\begin{align}\label{eqn:A}
 \left[i\partial_t+\omega_p\right] \destroy_\mu(2x) = &-i\tfrac{\vel}{2a} \left[\destroy_\mu(2x+a)-\destroy_\mu(2x-a)\right]
 \\
 \left[i\partial_t-\omega_p\right] \destroy_\mu(2x+a) = &-i\tfrac{\vel}{2a} \left[\destroy_\mu(2x+2a)-\destroy_\mu(2x)\right]
\end{align}
which combine to give
\begin{align}\label{eqn:KG_discretized}
\partial^2_t \destroy_\mu +\omega_p^2 \destroy_\mu = \tfrac{\vel^2}{4a^2} \left[\destroy_\mu(x+2a)+\destroy_\mu(x-2a)-2\destroy_\mu(x)\right].
\end{align}

Now, to limit the occupation number to 0 or 1 (to limit computational costs), we modify the Hamiltonian in Eq.~\eqref{eq:H_plasma-wave} to a model for hardcore bosons that has the same dispersion relation and then map the hardcore boson model to a spin model. 
The Hamiltonian that achieves this for hardcore bosons is
\begin{equation} \label{eqn:noninteracting bosons}
H = i\frac{\Ehop}{2}\sum_{j=1}^{N-1} (b_j\+ b_{j+1}\phant - b_{j+1}\+ b_j\phant) - \hbar\omega_p\sum_{j=1}^N (-1)^j b_j\+ b_j\phant.
\end{equation}
where $b_j \equiv \destroy_\mu(ja)$ and $\Ehop \equiv -2\hbar\vel/a$.
We then map the hardcore bosons to spins via
$b_j\+ \equiv \sigma^+_j$ and $b_j \equiv \sigma^-_j$,
which gives the spin Hamiltonian of Eq.~\eqref{eq:H}.
We note that the Heisenberg equations of motion for this model are not exactly those in Eq.~\eqref{eqn:KG_discretized}, but are modified to the version appropriate for hardcore bosons, which eliminates higher occupation numbers.

We believe that there is a path forward for extending this model to higher dimensions. In this case, one can use any representation of the Dirac gamma matrices with a complex Dirac spinor field. 
In dimensions $0-4 \mod 8$, the Majorana representation may be preferable because then the spinor field can be taken to be real. 
The sum over neighboring sites approximates the relevant Laplacian operators.
For example, for the case we study here, $ -\nabla\times\nabla=\nabla^2-\nabla\nabla\cdot$ is the perpendicular Laplacian for transverse EM waves and $\nabla\nabla\cdot $ is the parallel Laplacian for longitudinal Langmuir waves.

\section{Exact solution for the spin Hamiltonian}\label{app:exact_soln}
In Sec.~\ref{sec:model}, we considered the exactly solvable spin Hamiltonian
\begin{equation}
H = -\frac{\Ehop}{4}\sum_{i=1}^{N-1} (\sigma^x_i \sigma^y_{i+1} - \sigma^y_i \sigma^x_{i+1}) + \frac{1}{2}\sum_{i=1}^N \Egap_i (-1)^i \sigma^z_i.
\label{eq:H_app}
\end{equation}
We will assume uniform $\Egap_i = \Egap$. The standard way to solve this Hamiltonian is by mapping it to a spinless hardcore bosonic Hamiltonian~\footnote{Alternatively, one may map it to a free-fermionic Hamiltonian using a Jordan-Wigner transformation~\cite{jordan1993paulische}. They are equivalent in one dimension.},
\begin{align}
& \sigma^x_i = a_i\+ + a_i\phant\nonumber\\
& \sigma^y_i = i(a_i\+ - a_i\phant)\nonumber\\
& \sigma^z_i = a_i\phant a_i\+ - a_i\+a_i\phant,
\end{align}
where $a_i\phant (a_i\+)$ annihilates (creates) a hardcore boson at site $i$. On each site, the ground qubit state $\ket{0}$ maps to the vacuum of bosons, and the excited qubit state $\ket{1}$ maps to the singly occupied state. Under the hardcore-boson transformation, the spin Hamiltonian [Eq.~\eqref{eq:H_app}] maps to
\begin{equation}
H = i\frac{\Ehop}{2}\sum_{i=1}^{N-1} (a_i\+ a_{i+1}\phant - a_{i+1}\+ a_i\phant) - \Egap\sum_{i=1}^N (-1)^i a_i\+ a_i\phant.
\label{eq:H hardcore boson}
\end{equation}

After performing a Fourier transform on the lattice index, the hardcore boson Hamiltonian can be written as
\begin{equation}
H = \sum_k -\Ehop\sin (ka)\tilde{a}_k\+ \tilde{a}_k\phant - \Egap\sum_k \tilde{a}_k\+ \tilde{a}_{k+\pi/a}.
\label{eq:H free fermion k space}
\end{equation}
where $\tilde{a}_k\phant (\tilde{a}_k\+)$ annihilates (creates) a hardcore boson with wave vector $k$.
Two single-particle energy bands emerge, with energies
\begin{equation}
\hbar\omega_k = \pm \sqrt{\Ehop^2\sin^2ka + \Egap^2}.
\end{equation}
For periodic boundary conditions, the single-particle eigenstates are $\ket{\psi_k} = b_k\+ \ket{\textrm{vac}}$, with 
\begin{equation}
b_k = 
\frac{\left(\Ehop\sin ka - \hbar\omega_k \right) \tilde{a}_k + \Egap \tilde{a}_{k+\pi/a}}
{\sqrt{2\hbar\omega_k\left(\hbar\omega_k - \Ehop\sin ka\right)}}
.
\end{equation}
For open boundary conditions, the single-particle eigenstates are created by $b_k\+ = \sum_j c_{jk}\phant a_j\+ \ket{\textrm{vac}}$ where
\begin{align}
& c_{jk} = \sqrt{\frac{2}{N+1}}\sqrt{\frac{\hbar\omega_k+\Egap}{\hbar\omega_k}} \cos jka, & \textrm{if}\ j\ \textrm{is\ odd}, \nonumber
\\
& c_{jk} = -i\ \textrm{sgn}(\omega_k)\sqrt{\frac{2}{N+1}} \sqrt{\frac{\hbar\omega_k-\Egap}{\hbar\omega_k}} \sin jka, & \textrm{if}\ j\ \textrm{is\ even}.
\end{align}
Due to the hard-wall boundary condition, $ka = p \frac{\pi}{N+1}$ if $N$ is odd, and $ka = \left(p+\frac{1}{2}\right) \frac{\pi}{N+1}$ if $N$ is even, with $p$ an integer running from $0$ to $N/2$.
The allowed values for $ka$ cover only half of the usual range because it takes two coupled lattice sites to generate the two branches of the dispersion relation.
The physical interpretation is that the overall system size is really only $N/2$ because there are actually two coupled degrees of freedom per lattice site; i.e. the effective lattice spacing is actually $2a$. For example, for an EM plasma wave, the degrees of freedom at the two sites represent linear combinations of the charge density and electric current.

\section{Quantum Advantage}\label{app:advantage}
There has been recent development of so-called ``qubit lattice algorithms'' for simulating the classical physics of electromagnetic wave propagation in plasmas. Certain algorithms target classical computers \cite{vahala2020unitary, vahala2021one}, while others target quantum computers \cite{Koukoutsis2023dyson, Koukoutsis2023quantum}. While these works have obtained real speedups on classical supercomputers, and claim to offer efficient representations for quantum computers, they target the classical linearized equations of motion, where it has been much more difficult to find significant quantum advantage. While the recent work of Ref.~\cite{Babush2023prx} has determined classes of computationally intractable questions about the dynamics of linear oscillators that can be solved efficiently with quantum algorithms, it is more natural to investigate the potential speedup for intrinsically quantum dynamics.

Quantum computers are naturally adapted to simulating quantum systems and, thus, can offer significant quantum advantage for quantum dynamics. We approximated the quantum field theory of linear electromagnetic wave propagation in the low photon (or plasmon) density limit 
by a hard-core boson model that maps to a noninteracting fermionic model, which is an integrable model. While this makes classical simulations easy in certain cases, simulating the general case is still hard, as we argue in detail below.

The best classical algorithm for simulating a noninteracting quantum model depends on the initial condition.
Multiparticle states can be expressed using products of single-particle states, so simulating the evolution classically is tractable when the initial state can be written as a polynomial number of products of single-particle states. Let $M_0$ represent the number of single-particle states required to represent the initial condition. Then, one needs to simulate the evolution of each single-particle eigenstate in the multiparticle wave function efficiently.
For small enough problem sizes, one can use exact diagonalization, which generally scales as $\CO(M_0 N_s^3)$. If the overall number of single-particle states appearing in the initial condition is small or if the single-particle Hamiltonian matrix of size $\Ns\times\Ns$ is too large to form explicitly, then other methods, such as computing the matrix exponential for each state using only matrix-vector products would be cheaper \cite{moler2003nineteen, gaudreault2018kiops, bader2022skew-hermitian-exponential}.
Forming the matrix exponential for large problem sizes is often based on iterative Krylov-subspace methods \cite{gaudreault2018kiops} which have a cost scaling as $\CO(K N_s^2)$ where $K\leq N_s$ is the size of the largest Krylov subspace. 
However, if the number of single-particle states appearing in the initial condition, $M_0$, is large, then this cost becomes $\CO(M_0 N_s^3)$.
Thus, we conclude that the cost of the classical algorithm ranges from 
\begin{align}
    C\sim \CO(M_0 K N_s^2)\ \textrm{to}\ \CO(M_0N_s^3)
\end{align} depending on the initial conditions.
Computing the evolution of single-particle observables also has a similar cost, where $M_0$ now represents the number of independent observables.

In 1D, the Hamiltonian of interest is banded, and, in fact, is tridiagonal in our example. 
In this case, the worst scaling is not $\CO(\Ns^3)$ but rather $\CO(K N_s^2)$ where, in this context, $K$ represents the bandwidth or some measure of sparsity for a more general sparse matrix.

The initial state in the experiment that we implemented in Sec.~\ref{sec:plasma_wave_propagation} had one particle localized on two lattice sites. We only reported the density observable, which is a single-particle observable. The experiment was implemented in one dimension. Therefore, the time it takes to classically simulate one instance of that experiment is $C = \CO(N_s^2)$. This is, for example, why we were able to numerically simulate a large system with $N_s=100$, as shown in Fig.~\ref{fig:large}.

Even for these simple cases, using a quantum computer to simulate the quantum problem, as explored in this work, provides a complexity advantage. The quantum algorithm only requires simulating the problem in time and does not require performing an eigendecomposition. The cost of Trotterizing the evolution operator scales as the number of interactions between sites, $\nint$, and, because the Hamiltonian only has local interactions, this scales as the spatial dimension $\nint\propto\ddim$. Thus, evolving this quantum algorithm, which scales as $\nint$ per time step, for the same total time interval only has a cost of $\CO( \nint\Nt)$. Due to the Courant-Friedrichs-Lewy (CFL) criterion for stability and accuracy, the number of time steps scales linearly in grid spacing $\Nt\propto\ddim \Nx$. 
Therefore, the total time for the quantum algorithm scales as 
\begin{align}
    Q = \CO(\ddim^2 \Nx).
\end{align}
Thus, the quantum speedup ranges from 
\begin{align}
C/Q=\CO(M_0 K \Nx^{2\ddim-1}/\ddim^2) 
\, \text{--}\,
\CO(M_0 \Nx^{3\ddim-1}/\ddim^2).
\end{align}

Nonetheless, in the most general case, the initial state may be any state that is easy to prepare with a quantum circuit but hard to express in terms of the noninteracting model, and the observable of interest may be any easily measurable observable in the quantum circuit but highly nonlocal or multisite in the particles.
As a concrete example, a state that is easy to prepare with a quantum circuit but hard to express in terms of the noninteracting model, is the equal superposition of all computational basis states, which can be prepared by applying a Hadamard gate to all the qubits. 
In such cases, the only way to classically compute the observables of interest is via exact diagonalization of the full Hilbert space whose size is $2^{N_s}$, which takes exponential time. 
In these cases, executing the quantum circuit on a quantum computer has an exponential advantage over classically simulating the circuit. As such, exactly classically simulating a circuit with $>50$ qubits is typically out of reach of current classical computers. We note that the best classical methods to simulate quantum systems in 1D are time-dependent matrix-product-state (MPS) methods, but even these would struggle due to the need for a rapidly increasing bond dimension with time. Moreover, MPSs scale poorly in $>1D$, and higher-dimensional tensor-network methods are difficult to compute efficiently.

Finally, although the quantum algorithm still has a scaling advantage for the simple cases we considered in this paper, it does not necessarily translate to a practical advantage when real time scales are compared at current qubit scales. Indeed, for the model we studied, quantum advantage is not expected even for $\Ns = \CO(10^5)$ qubits in one dimension. If we consider a more general Hamiltonian, e.g. with additional ZZ interactions, then it does not map to a noninteracting model, and this model cannot be classically simulated in $\poly(\Ns)$ time. Solving this system classically generally takes exponential time in $\Ns$, where $C\sim \CO(2^{3\Ns})$.

\section{Initial conditions}
For classical plasma simulation, one would need to specify two initial conditions, $\VA(\Vx)$ and $\VE(\Vx)$, at the initial time, as well as boundary conditions for the given time interval.
However, because the fields are canonically conjugate, for the quantized problem, the fields do not commute and must instead obey the Heisenberg uncertainty relations.
Thus, one cannot specify both or measure both fields at the same time and must set the conditions for the wave function or density matrix in a manner consistent with the laws of quantum mechanics.
This can be performed by setting boundary conditions for the spin Hamiltonian and setting an initial spin-wave function or density matrix.
For example, one could specify a definite initial spin orientation or one could specify the initial conditions in terms of spin coherent states, which balance the uncertainty in the fields in an optimal manner.
To compare this to a classical simulation, observables such as the energy of any given configuration can be determined by their expectation value.

\section{Hardware characteristics}\label{app:hardware}
The Rigetti Ankaa-3 quantum processing unit (QPU) is composed of transmon qubits connected via floating tunable couplers (which are also transmons) \cite{sete_floating_2021} arranged in a square grid. The tunable couplers enable control of the effective qubit-qubit coupling between neighboring qubits via the modulation of their frequency, which is controlled by threading external magnetic flux through the superconducting quantum interference device (SQUID) loops. The coupling is turned on by applying a baseband flux pulse to the coupler when actuating a two-qubit gate.
The use of tunable couplers allows the interaction strength between the two qubits to be tuned, minimizing unwanted interactions with spectator qubits during the operation of gates while enhancing the interaction strength between the partners of the two-qubit gate.

Entangling gates between neighboring qubits are implemented using a bipolar baseband flux pulse applied to the higher-frequency qubit, bringing it into resonance with its neighbor \cite{Sete_para_2021}. This induces an XX+YY interaction, typically accompanied by a small ZZ component. In the frame rotating at the qubit frequencies, this interaction realizes the so-called FSIM gate~\cite{rosenberg2024dynamics} up to one-qubit phases. We define the FSIM gate as
\begin{align}
&{\rm FSIM}(\theta,\phi) = \nonumber\\
&\left( \begin{array}{cccc}
1 & 0 & 0 & 0\\
0 & \cos\frac{\theta}{2} & i \sin\frac{\theta}{2} & 0\\
0 & i \sin\frac{\theta}{2} & \cos\frac{\theta}{2} & 0\\
0 & 0 & 0 & e^{i\phi}
\end{array}\right).
\end{align}
We note that this definition has the opposite signs for $\theta$ and $\phi$ as compared to some other conventions, e.g., that given in Ref.~\cite{cirq_fsim}. The native gate realized on the device is a ``PHASEDFSIM'' gate,
\begin{align}
&{\rm PHASEDFSIM}(\theta,\phi,\zeta,\gamma,\chi) = \nonumber\\
&\left( \begin{array}{cccc}
1 & 0 & 0 & 0\\
0 & e^{-i(\gamma+\zeta)}\cos\frac{\theta}{2} & i e^{-i(\gamma-
\chi)}\sin\frac{\theta}{2} & 0\\
0 & i e^{-i(\gamma+\chi)}\sin\frac{\theta}{2} & e^{-i(\gamma-\zeta)}\cos\frac{\theta}{2} & 0\\
0 & 0 & 0 & e^{i(\phi-2\gamma)}
\end{array}\right),
\end{align}
which is equal to $Rz\left(\frac{\zeta-\chi}{2}-\gamma\right)\otimes Rz\left(\frac{\chi-\zeta}{2}-\gamma\right) \cdot {\rm FSIM}(\theta,\phi) \cdot Rz\left(\frac{\zeta+\chi}{2}\right)\otimes Rz\left(-\frac{\zeta+\chi}{2}\right)$.
Here, $\theta$ and $\phi$ are iSWAP-like and ZZ-like entangling phases, respectively, and the corresponding iSWAP-like and ZZ-like interactions are also referred to as transverse and longitudinal interactions. For example, FSIM$(\theta=\pi,\phi=0)$ is the iSWAP gate, and FSIM$(\theta=0,\phi=\pi)$ is the CZ gate. Here, $(\chi,\zeta,\gamma)$ are single-qubit phases that arise from the qubits' different idle frequencies and their frequency excursions during the two-qubit gate. The $\chi$ phase depends on the time at which the gate is played in the circuit, and it is given by $\chi = \Delta\nu t$, where $\Delta\nu$ is the difference in the qubits' idle frequencies and $t$ is the time difference between the start of the circuit and the time at which this gate is executed. We do not assume this form for $\chi$ and instead learn $\chi$ along with the other angles using the gate-learning technique described in Sec.~\ref{subsec:unitary_learning}.

We tuned the gates such that $\theta\simeq \pi/2$ and $\phi\simeq0$ on each edge, and more accurately inferred $(\theta,\phi,\zeta,\chi,\gamma)$ for each edge using tomography and a variant of cross-entropy benchmarking (see Sec.~\ref{subsec:unitary_learning}).

The characteristics of the Ankaa-3 device at the time of the experiment are shown in Table~\ref{tab:ankaa-3-char}. The T1 and T2 values are reported on Rigetti Quantum Cloud Services while the gate fidelities were measured directly prior to the experiment. For the experimental configuration of circuits executed in this report, the EPLG was measured to be $1.48 \pm 0.02\%$.

\section{Executing the experiment}
\begin{figure*}[t]
\centering
\includegraphics[width=0.7\textwidth]{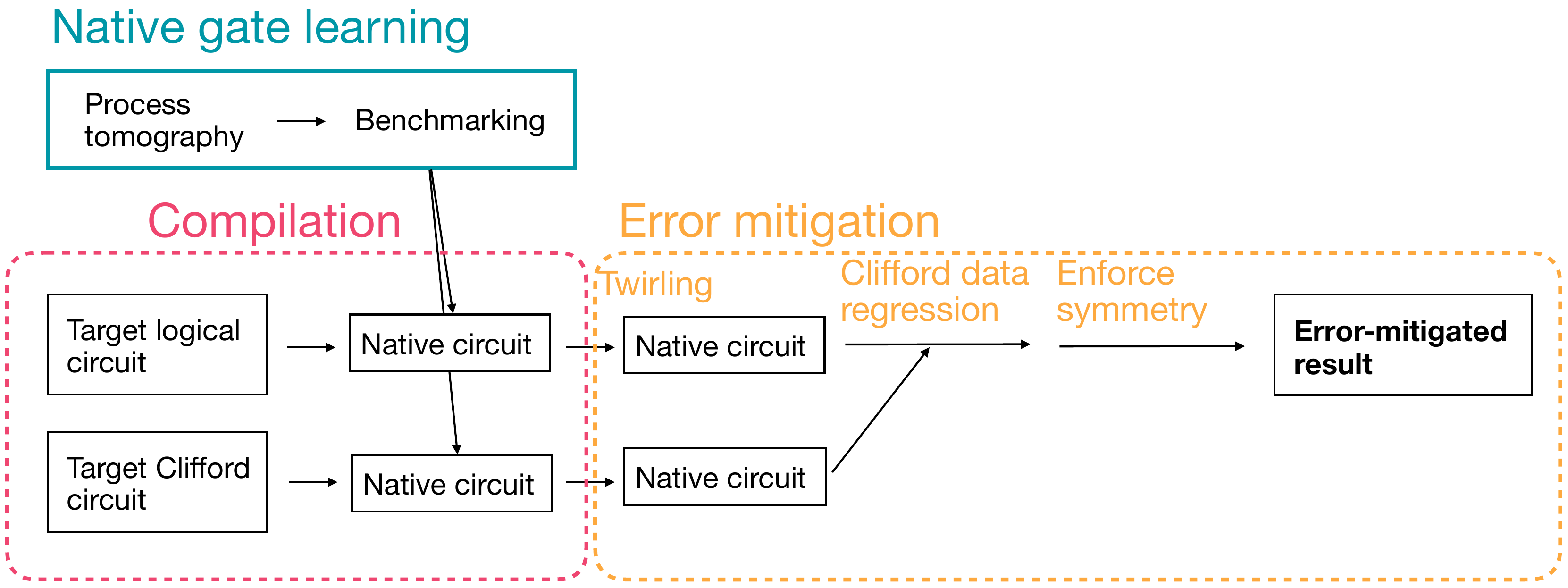}
\caption{{\bf A schematic overview of the experiment.} A two-step gate-learning protocol learns the native two-qubit gates on the hardware. These native gates are used in the compilation of the target logical circuit. Meanwhile, a set of Clifford circuits with the same gate placements as the target circuits are also executed. The results of the target circuits are subject to error mitigation using Clifford data regression.}
\label{fig:sm-overview}
\end{figure*}

A schematic overview of the experimental pipeline is depicted in Fig.~\ref{fig:sm-overview}.
Before executing the logical target experiments, we perform a few steps. First, we learn the native entangling gates between the qubits, as explained in Sec.~\ref{subsec:unitary_learning}. Then, we compile the target logical circuit into logical gates between the qubits (Sec.~\ref{subsec:trotter}) and express the logical gates in terms of native gates (Sec.~\ref{subsec:numerical_compilation}). Once we have all the circuit instances ready for execution, we construct a list of Clifford circuits with the same brickwork gate structure as the target circuits. The Clifford circuits are useful for error mitigation, as explained in Sec.~\ref{subsec:cdr}. We transform all the logical circuit instances and Clifford circuit instances using twirling (Sec.~\ref{subsec:twirling_fsim}). Finally, we execute all our circuits and take shots, implement error mitigation (Appendix~\ref{app:error_mitigation}), and calculate the error-mitigated observables.

\subsection{Gate-learning via tomography and benchmarking}\label{subsec:unitary_learning}

An important step in executing our experiments is learning the gates between the qubits. We learn the native gate between the qubits in a two-step process: first, a rough estimation of the gate using process tomography, and then a finer estimation of the gate using a 
variant of randomized benchmarking.

Process tomography is a technique that is commonly used to learn a quantum process. The basic recipe is to prepare a complete set of basis states, apply the process, and measure a complete set of measurement bases~\cite{mohseni2008quantum}. 
To learn the gate from the process-tomography results, the superoperator is first reconstructed from the observable data. This includes the effects of noise, but the condition that the superoperator be physical, i.e. completely positive and trace preserving, is enforced. We numerically maximize the fidelity of this superoperator against a parameterized PHASEDFSIM candidate. This gives us a rough calibration of the native entangling gate. The process tomography was performed on an entire entangling gate cycle at once, allowing the unitaries to be learned in context.

\begin{figure}[t]
\centering
\includegraphics[width=0.8\columnwidth]{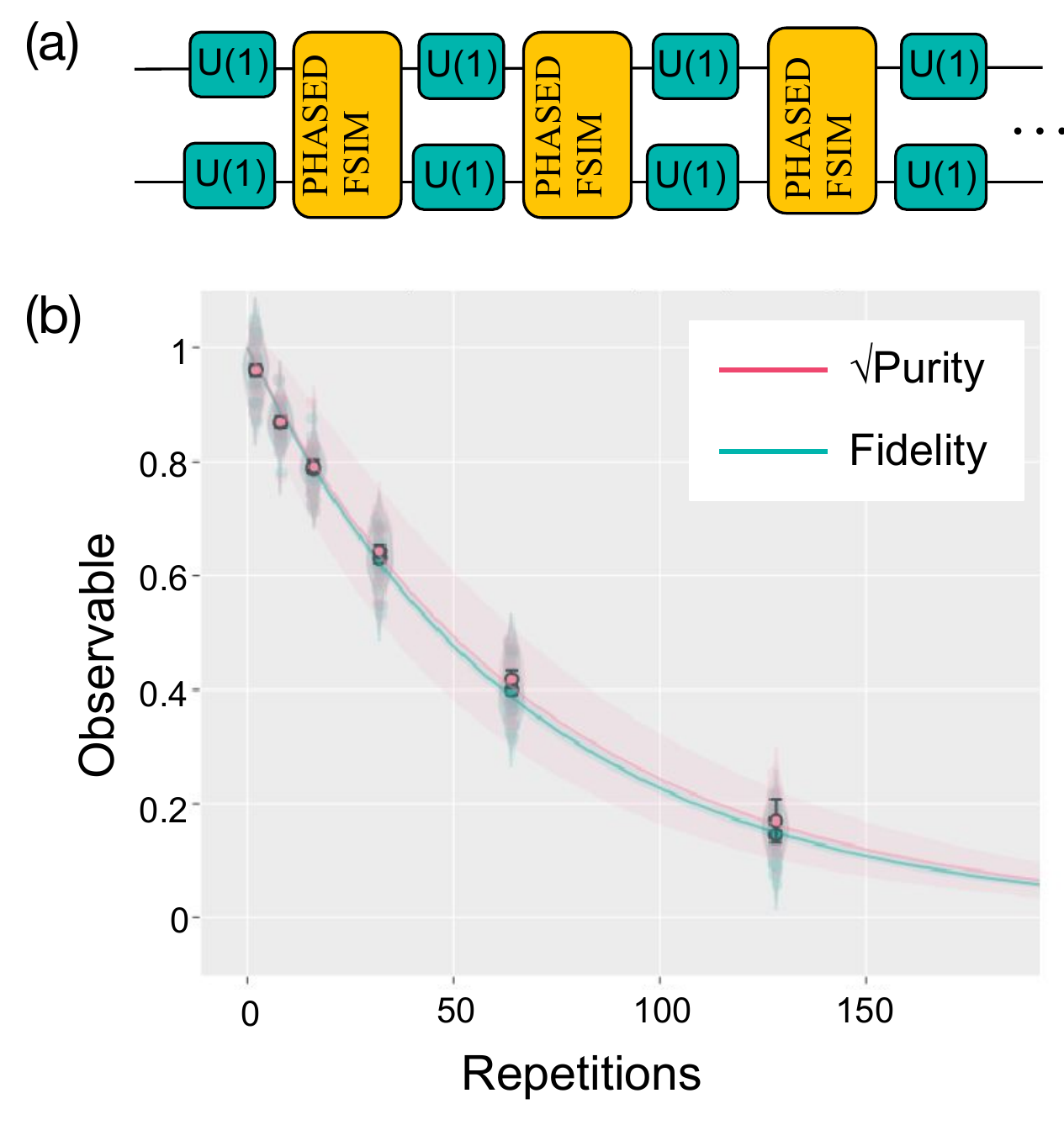}
\caption{{\bf Schematic of a shadow benchmarking experiment}. (a) A circuit with a long sequence of PHASEDFSIM gates interleaved with random one-qubit gates is implemented. (b) The decay of the shifted fidelity (green) of the experimentally prepared state with a noiseless numerical simulation, and the square root of the shifted purity (red) of the state in the experiment, versus sequence length. A classical routine learns the gate parameters such that the fidelity is maximized. The gap between the shifted fidelity and purity curves can indicate a unitary error, which is very small here. The benchmarking is performed in context.}
\label{fig:fsim-shadow-benchmarking}
\end{figure}

To obtain a finer calibration of the entangling gate, we use a method inspired by cross-entropy benchmarking~\cite{arute2019quantum}. A circuit is constructed with many layers of the two-qubit gate cycle interleaved with random one-qubit gates, as shown in Fig~\ref{fig:fsim-shadow-benchmarking}(a). Then, we calculate the fidelity of the output state with the predicted state in a noiseless numerical simulation for the circuit. Similar to above, we numerically maximize the fidelity of the experimentally produced state against the state produced by a parameterized PHASEDFSIM candidate. The $\chi$ phase of each gate in the circuit is parameterized with two parameters $(\chi_0,\Delta\nu)$ as $\chi = \chi_0 + \Delta\nu t$, where $t$ is the time at which the gate appears in the circuit. The other four parameters are constant for all the PHASEDFSIM gates on the same edge in the circuit. Learning the gate parameters from a circuit with many layers of the two-qubit gate provides robustness to state preparation and measurement (SPAM) errors.

Because the benchmark circuit consists of separable states with independently entangled pairs of qubits, this technique is scalable. We note that unlike cross-entropy benchmarking, where the fidelity is estimated using the cross-entropy of the observed bitstrings with the expected bitstring distribution, we instead estimate the state fidelity using direct fidelity estimation, applying the approach of classical shadows \cite{huang2020predicting}. The main reason to use classical shadows is to mitigate nonMarkovian errors in the measurements of the qubits.

Since the benchmarking sequence consists of a long sequence of gates, the results will be affected by noise. Typically, the purity $\mathcal{P}$ and the fidelity $\mathcal{F}$ (with the ideal state computed from the right gate parameters) of the states decay from $1$ to $1/4$, as the sequence length approaches infinity. It is therefore useful to compute the shifted fidelity and the shifted purity,
\begin{align}
\tilde{\mathcal{P}} = \frac{4\mathcal{P}-1}{3} \nonumber\\
\tilde{\mathcal{F}} = \frac{4\mathcal{F}-1}{3}
\end{align}
at each sequence length. For a depolarizing noise model, the shifted purity and fidelity would decay as $f^{2n}$ and $f^n$, respectively. Here, $f$ is known as the unitarity of the noise channel. Any difference between the shifted fidelity and the square root of the shifted purity is indicative of coherent errors.

To perform direct fidelity estimation at each sequence length, we use the classical shadow to estimate all $(4^2-1)$ Pauli observables for each two-qubit subsystem. The fidelity is given by
\begin{equation}
\mathcal{F} = \frac{1}{2^N}\sum_P \braket{P}_{\textrm{experiment}} \braket{P}_{\textrm{ideal}}
\end{equation}
where $N=2$ qubits here.
The uncertainty of each Pauli observable estimate is determined by its Pauli weight $w$ and the sample count $M$, and it is given by $\sim\sqrt{3^w / M}$. A typical state produced by the benchmarking circuit is Haar-random; therefore, $\braket{P}_{\textrm{ideal}} \sim 1/\sqrt{2^N}$. Putting these together, the uncertainty $\Delta F$ of the fidelity is given by
\begin{equation}
(\Delta \mathcal{F})^2 = \frac{1}{4^N}\sum_{w=0}^N \begin{pmatrix}N\\w\end{pmatrix} 3^w \frac{3^w}{M} \frac{1}{2^N}
\end{equation}
since there are $\begin{pmatrix}N\\w\end{pmatrix} 3^w$ Paulis with weight $w$. Using the binomial theorem, we simplify the above equation to $(\Delta \mathcal{F})^2 = \frac{1}{M}\left(\frac{5}{4}\right)^N$. For $N=2$ and $M= 3,000$ shots, we can thus estimate the fidelity of the final two-qubit state to an uncertainty of $\Delta \mathcal{F} = 2.3\%$. For comparison, the cross-entropy uncertainty, $\Delta \hat{H}_{lin} = 1 / \sqrt{M}$. The cross-entropy metric does not depend on the system size, which is a useful property for scalable benchmarking, but in the regime of $N=2$, the direct fidelity method requires only 25$\%$ more samples and has the advantage of being bounded between 0 and 1.

Our experiment also allows measurement of the purity of the state,
\begin{equation}
\mathcal{P} = \frac{1}{2^N}\sum_P \braket{P}_{\textrm{experiment}}^2,
\end{equation}
which we will use later to track coherent errors in the gate. A similar calculation to above, and using $\Delta(\braket{P})^2 = 4|\braket{P}|^2(\Delta P)^2$, yields
\begin{equation}
(\Delta \mathcal{P})^2 = \frac{1}{4^N}\sum_{w=0}^N \begin{pmatrix}N\\w\end{pmatrix} 3^w 4\times \frac{3^w}{M} \frac{1}{2^N} = \frac{4}{M}\left(\frac{5}{4}\right)^N.
\end{equation}
For $N=2$ and $M= 3,000$ shots, we can thus estimate the purity of the final two-qubit state to an uncertainty of $\Delta \mathcal{P} = 4\%$.

In our benchmarking experiments, we typically average these estimates over 30 random sequences. An example decay with a fit is shown in Fig.~\ref{fig:fsim-shadow-benchmarking}(b). The use of effective readout-error mitigation \cite{arrasmith2023development} with reasonable assumptions about the noise allows us to assert that the estimate begins at 1 and decays to 0, and we use the simplified fit form of $f(n) = f^{n}$, allowing for fewer points and higher confidence in the fidelity estimate.

\subsection{Trotterized evolution}\label{subsec:trotter}
\begin{figure}[t]
\centering
\includegraphics[width=0.7\columnwidth]{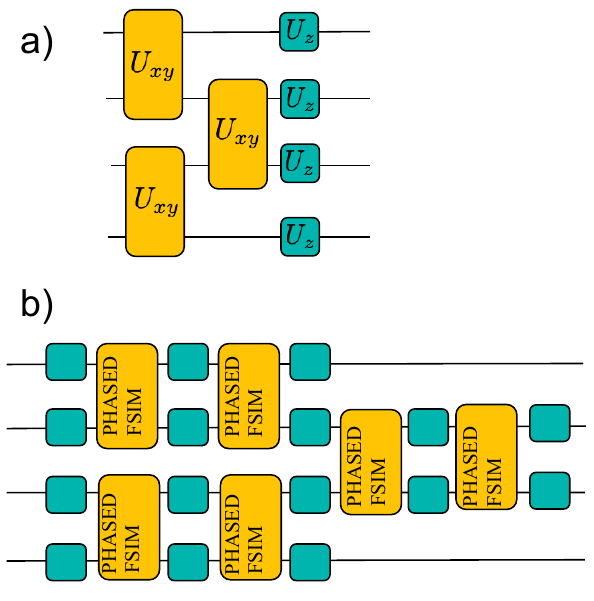}
\caption{{\bf Circuit for Trotterized evolution.} (a) A single Trotter evolution step with the Hamiltonian $H$ [Eq.~\eqref{eq:H}]. Here, $U_{xy}$ and $U_z$ are the evolutions with the spin-spin interaction and plasma-gap terms in the Hamiltonian, respectively. b) Approximate decomposition of one Trotter evolution step into native two-qubit gates and one-qubit gates implemented via the PMW4 scheme.}
\label{fig:sm-trotter}
\end{figure}

We implement evolution with $H$ using a first-order Trotter expansion. Each Trotter step can be conceptually split into three logical layers -- one layer implements evolution with $\sigma^x_i\sigma^y_{i+1} - \sigma^y_i\sigma^x_{i+1}$ for odd $i$, one layer implements the above for even $i$, and one layer implements evolution with the $\sigma^z_i$ terms in $H$. We denote the evolution due to the interaction terms as $U_{xy}$ in Fig.~\ref{fig:sm-trotter}(a), and evolution due to the $\sigma^z$ terms as $U_z$. These logical layers are then compiled to native gates using an approximate numerical compilation explained in Sec.~\ref{subsec:numerical_compilation}. In practice, the $U_z$ gates are not implemented separately from the two-qubit layers; They are absorbed into the compilation of the two-qubit layers. Each compiled Trotter step has the structure shown in Fig.~\ref{fig:sm-trotter}(b).

\subsection{PMW4 decomposition}\label{subsec: pmw4}
Rigetti Ankaa-class devices can realize $\text{RX}(\pi)$ and $\text{RX}(\pm\pi/2)$ gates, where $\text{RX}(\theta) = \exp(-i\frac{\theta}{2}\sigma^x)$. Ankaa-class devices also realize a parametric $\text{RZ}(\theta) = \exp(-i\frac{\theta}{2}\sigma^z)$ gate using local updates of in-sequence phases that consume zero runtime and introduce negligible error \cite{mckay_efficient_2017}. Any one-qubit operation can be implemented using at most three $\text{RZ}(.)$ and two $\text{RX}(\pi/2)$ operations, where each $\text{RX}(\pi/2)$ is implemented via a pulse, and each $\text{RZ}(.)$ determines the in-sequence phase update~\cite{barenco1995elementary}. The above implementation of the RZ gate is called a virtual RZ, and it is useful when two-qubit gates $U$ in circuits are phase-carrier gates, i.e. they satisfy
\begin{equation}
U \cdot(\text{RZ}(\theta_1) \otimes \text{RZ}(\theta_2)) = (\text{RZ}(\theta_3) \otimes \text{RZ}(\theta_4)) \cdot U
\end{equation}
for some $\theta_{i=1\cdots 4}$.
In this case, the $\text{RZ}$ gate is just carried over the two-qubit phase-carrying gate, and the in-sequence phases are adjusted accordingly.

The PHASEDFSIM gate, however, is not phase carrying; therefore, the virtual RZ scheme is not feasible with PHASEDFSIM gates. Instead of the above scheme, we thus implement one-qubit gates using four microwave pulses, known as the PMW4 scheme~\cite{chen2023compiling},
\begin{equation}
U_{1Q} = X_{\pi/2}(\theta)X_{\pi/2}(\phi)X_{\pi/2}(\phi)X_{\pi/2}(\omega)
\end{equation}
where $X_\alpha(\phi) \equiv \text{RZ}(-\phi)\text{RX}(\alpha)\text{RZ}(\phi)$. In this scheme, the net phase advanced by a one-qubit gate is 0; therefore, no phase needs to be carried forward by the PHASEDFSIM gate.

\subsection{Approximate numerical compilation of gates}\label{subsec:numerical_compilation}
Once we have calibrated arbitrary one-qubit and native two-qubit gates, we are ready to implement arbitrary logical circuits. Implementing an arbitrary logical two-qubit gate $U_{\textrm{tgt}}$ requires us to express it in terms of the native gates. We do this using an approximate numerical compilation technique.

Our compilation scheme finds the best compilation that uses at most two PHASEDFSIM gates to express the target logical gate; i.e., we express
\begin{align}
U_{\textrm{approx}} = & (u_1 \otimes u_2)\cdot \textrm{PHASEDFSIM} \cdot (u_3 \otimes u_4) \cdot \nonumber\\ 
& \textrm{PHASEDFSIM}(u_5 \otimes u_6)
\label{eq:approx_compile}
\end{align}
where $\textrm{PHASEDFSIM}$ refers to the learned native $\textrm{PHASEDFSIM}(\theta,\phi,\zeta,\chi,\gamma)$ on the edge. Here, $u_i$ are parameterized single-qubit gates, where we numerically find the parameters such that $|\textrm{tr}(U_{\textrm{approx}}\+ U_{\textrm{tgt}})|$ is maximized, i.e. $U_{\textrm{approx}}$ is as close to $U_{\textrm{tgt}}$ as possible.
We point out that Eq.~\eqref{eq:approx_compile} only has two PHASEDFSIM gates, meaning that it is not possible to exactly express the full range of two-qubit gates. 
Nonetheless, it can be useful to approximately express gates \cite{jurcevic_demonstration_2021}, provided that the error in doing so is smaller than the error incurred by an additional native entangling gate. Thus, the approximate-expression technique has broad applicability. For the target logical gates in our experiment, the maximum compilation error $1-|\textrm{tr}(U_{\textrm{approx}}\+ U_{\textrm{tgt}})|^2/16$ had a relatively small value of 0.12\%.

The numerical compilation scheme involves a classical optimization of the $u_i$, which is expensive and can become a bottleneck. Therefore, to enable fast compilation, we developed a vector database of expression instances. An expression instance is composed of both the target unitary $U_{\textrm{tgt}}$, and the native entangling unitaries $\textrm{PHASEDFSIM}(\theta,\phi,\zeta,\chi,\gamma)$ which attempt to express it when combined with the arbitrary single-qubit rotations. The goal is to retrieve the value of the expression instance from the database; however, instances rarely match to floating-point precision and unitaries are equivalent up to a global phase, making traditional databases unsuitable for storing instances. Therefore, we turn to a vector database, where the keys are vectors and the retrieval is based on a distance metric \cite{noauthor_chroma-corechroma_2025}.

We formulate the instance matching as the minimization of a distance metric. The process fidelity of two unitary matrices $U$ and $U'$, $F(U, U') = \left(\frac{{\rm Tr}[U\+ U']}{4}\right)^2 = \frac{{\rm Tr}[\mathcal{U}\+ \mathcal{U}']}{16}$, where $\mathcal{U}$ and $\mathcal{U}'$ are the Pauli-Liouville matrices for $U$ and $U'$, is maximized when $U = U'$ up to a global phase. We note that the trace can be written as ${\rm Tr}[\mathcal{U}\+ \mathcal{U}'] = \vec{\mathcal{U}}\cdot \vec{\mathcal{U}}'$, where $\vec{\mathcal{U}}$ and $\vec{\mathcal{U}}'$ are vectors obtained by raveling the superoperators. Thus, we can define a \textit{distance} between two unitaries as $d(U,U') = |\vec{\mathcal{U}}\cdot \vec{\mathcal{U}}' - 1|$. To find the closest match of an expression instance $\left(U_{\textrm{tgt}}, \textrm{PHASEDFSIM}, \textrm{PHASEDFSIM}\right)$ with instances $(u, \textrm{PHASEDFSIM}, \textrm{PHASEDFSIM})$ in the database, we minimize the sum of the distances, $d(U_{\textrm{tgt}},u) + 2d(\textrm{PHASEDFSIM},\textrm{PHASEDFSIM})$.

\section{Error mitigation}\label{app:error_mitigation}
Quantum hardware suffers from several kinds of noise that introduce errors into their output. Error mitigation has been shown to be a useful technique for evaluating observables of interest in noisy quantum computers~\cite{kim2023evidence}. Generally, one first tailors the hardware noise to a form that is amenable to mitigation. Then, an appropriate error-mitigation technique is implemented. Various error-mitigation techniques have been proposed in the literature~\cite{temme2017error}. Here we describe our noise-tailoring and error-mitigation methods.

\subsection{Twirling}\label{subsec:twirling_fsim}
Noise in gates is often described via quantum channels. For example, the action of a noisy gate on a state $\rho$ is written as $\mathcal{U}_{\textrm{noisy}}(\rho) = \Phi \mathcal{U} \rho$, where $\mathcal{U}$ is the noiseless gate $U$ in superoperator space, and $\Phi$ is the noise channel. We use calligraphic symbols for operators in superoperator space, normal font for operators in Hilbert space, and capital Greek letters for noise channels.

Typically, the noise channel $\Phi$ on the device is not known. It is generally desirable to minimize the coherent noise, since coherent errors can interfere constructively.

The most common method for transforming coherent noise into incoherent noise is Pauli twirling. Pauli twirling converts arbitrary noise to stochastic Pauli noise, which is more amenable to error mitigation than other noise. In the above case for measuring unitarity, for example, the experiment would have layers of the target two-qubit gate interleaved with random Pauli gates, which would convert the coherent noise to incoherent noise.

\subsubsection{Background: Pauli twirling}
Twirling maps a circuit to a logically equivalent circuit if the gates were noiseless. Pauli twirling does this by injecting random Pauli gates and their conjugates into the circuit. Any Clifford gate $U$ may be equivalently written as $P U P'$, where $P$ and $P'$ are conjugated Pauli products. An ensemble of circuits is executed, where the circuits twirl $U$ with different Paulis $P$, and the ensemble average for a target observable is computed. Averaging over all Pauli twirls, the noisy gate $\mathcal{U}_{\textrm{noisy}}$ gets transformed to 
\begin{equation}
\mathcal{U}_{\textrm{noisy}} \rightarrow \frac{1}{16}\sum_\mathcal{P} \mathcal{P} \mathcal{U}_{\textrm{noisy}} \mathcal{P}' \equiv \frac{1}{16}\sum_\mathcal{P} \mathcal{P} \Phi \mathcal{U} \mathcal{P}'
\label{eqn:Pauli_twirl}
\end{equation}
where the factor $16$ in the denominator refers to the number of two-qubit Pauli products. Since $\mathcal{P} \mathcal{U} \mathcal{P}' = \mathcal{U}$, Eq.~\eqref{eqn:Pauli_twirl} becomes $\mathcal{U}_{\textrm{noisy}} \rightarrow \Phi_{\textrm{twirl}} \mathcal{U}$ with $\Phi_{\textrm{twirl}} = \frac{1}{16}\sum_\mathcal{P} \mathcal{P} \Phi \mathcal{P}\+$. It can be shown the twirled noise $\Phi_{\textrm{twirl}}$ is stochastic Pauli noise.

\subsubsection{Twirling the FSIM gate}
The native entangling gate used in our experiment, the FSIM gate, is not a Clifford gate, and therefore it cannot be twirled by Pauli gates; however, we show below that it is possible to partially twirl the FSIM gate using only one-qubit gates. Since PHASEDFSIM$(\theta,\phi,\zeta,\chi,\gamma)$ is equal to FSIM$(\theta,\phi)$ sandwiched between additional $Rz$ gates, it is then straightforward to implement the one-qubit gates that twirl PHASEDFSIM$(\theta,\phi,\zeta,\chi,\gamma)$ which are the actual native gates on our device. 

\begin{figure}[t]
\centering
\includegraphics[width=0.7\columnwidth]{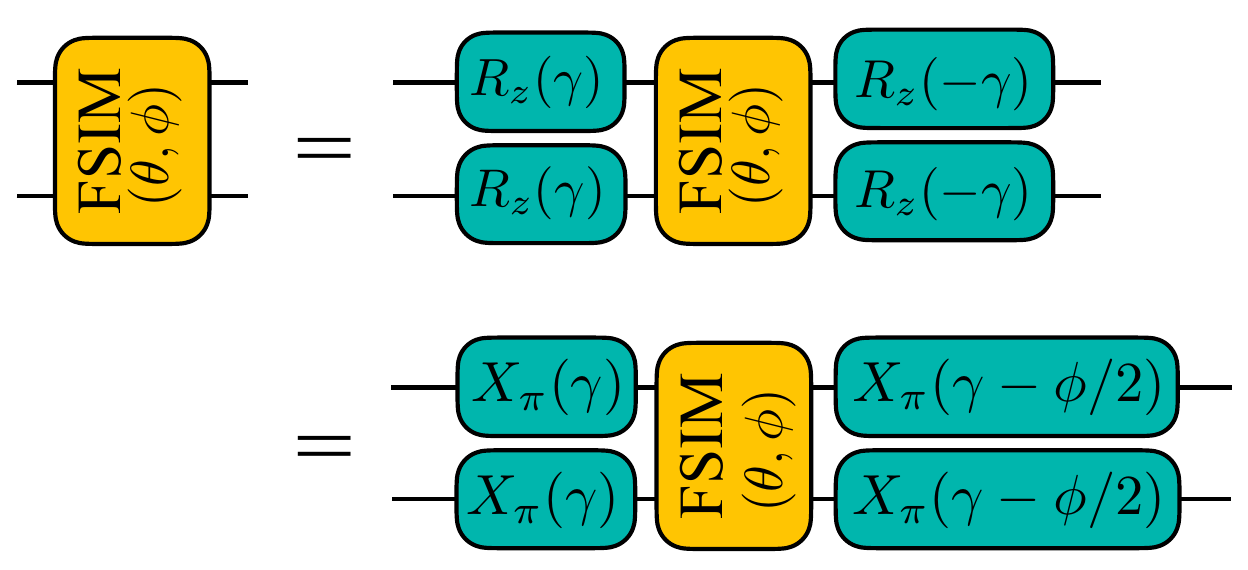}
\caption{Two continuous families of the FSIM twirling group, which are functions of the continuous parameter $\gamma$. The symbol $X_\pi(\gamma)$ means $R_z(-\gamma)R_x(\pi)R_z(\gamma)$ (defined earlier in Sec.~\ref{subsec: pmw4} and repeated here for convenience). Three distinct Pauli twirls can be obtained as special cases of these continuous families. For example, the ZZ twirl is obtained by setting $\gamma = 0$ in the first line, the XX twirl by setting $\gamma = 0$ in the second line, and the YY twirl by setting $\gamma = \pi/2$ in the first line. 
Due to the absence of the remaining 12 Pauli twirls, errors are only partially twirled by this group.}
\label{fig:sm-fsim-twirling-group}
\end{figure}

For the FSIM$(\theta,\phi)$ gate, a subset of the Pauli twirls can be efficiently inverted, specifically the II, XX, YY, and ZZ twirls. Furthermore, we found two continuous families of pairs of one-qubit gates that twirl the FSIM$(\theta,\phi)$ gate, as shown in Fig.~\ref{fig:sm-fsim-twirling-group}. We note that setting $\gamma=\pi$ in the first family of twirls (first line in Fig.~\ref{fig:sm-fsim-twirling-group}) gives the twirl with ZZ, setting $\gamma=0$ in the second family of twirls (second line in Fig.~\ref{fig:sm-fsim-twirling-group}) gives the twirls with XX, and $\gamma = \pi/2$ in the second line gives the twirls with YY. Furthermore, the entire family of twirls in the second line at arbitrary $\gamma$ can be obtained by composing the first line and twirling with XX.

Unlike the Paulis, these twirls are unable to fully convert an arbitrary coherent error channel into a stochastic one, for some sensible choice of distribution over $\gamma$. Rather, certain coherent errors are twirled, while others are not. Notably, over-rotation-type errors of the FSIM gate, where the angles $\theta$ or $\phi$ are different from their calibrated value, are not twirled (see also Ref.~\cite{tsubouchi2024symmetric}). We aim to learn the gate such that over-rotation errors, which could occur due to errors in calibration, are negligible. We assume the angles to be stable over the course of the experiment, however over a long enough time scale, drift can result in the calibration becoming stale.
Nevertheless, the process learns only unitaries of the FSIM type, and it thus cannot capture other classes of coherent errors. The twirling group is well suited to addressing this type of coherent error, which may arise from second-order interactions and control errors.

To understand the effect of the twirling group, we turn to some examples and examine the unitary and nonunitary components of the twirled superoperator. In what follows, the overall infidelity is 
$$\mathcal{E}_F = 1 - \text{Tr}(\Phi) / d^2$$
where $\Phi$ is the superoperator of the channel, $d$ is the dimension, and the stochastic, or incoherent, infidelity is
$$\mathcal{E}_S = 1 - \sqrt{\text{Tr}{(\Phi \Phi^{\dagger})}} / d$$
The coherent infidelity is then the overall infidelity minus the stochastic infidelity,
$$\mathcal{E}_U \equiv \mathcal{E}_F - \mathcal{E}_S = \frac{ \sqrt{\text{Tr}{(\Phi \Phi^{\dagger})}} }{ d } - \frac{ \text{Tr}(\Phi) }{ d^2 }.$$
Twirling preserves the trace of $\Phi$; i.e., it does not change the overall infidelity, but converts some or all of the coherent infidelity to incoherent infidelity.

\begin{figure}[t]
\centering
\includegraphics[width=1.0\columnwidth]{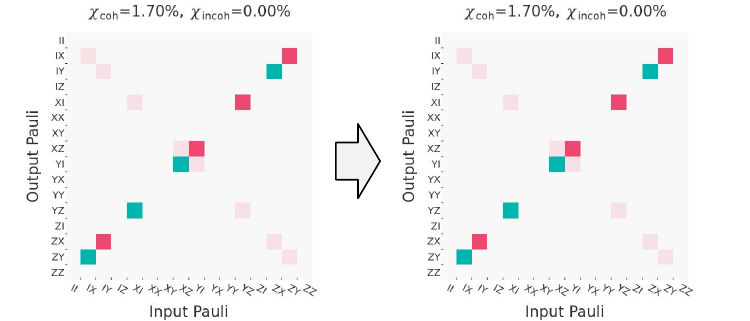}
\caption{The FSIM twirling group is applied to a RZZ($\pi$/12) error. The two panels show $\Lambda-1$ for the initial and twirled error channels, where $\Lambda$ is the Pauli transfer matrix. Left : the initial error channel has an overall infidelity of $\sin^2(\pi/24) \approx 1.70\%$ entirely due to coherent error. Right: the twirled channel, which also has a coherent infidelity of 1.70\%. Thus, the RZZ-type error is not tailored at all by the twirling group. This type of error must be addressed by FSIM learning.}
\label{fig:rzz-twirl}
\end{figure}

\begin{figure}[t]
\centering
\includegraphics[width=1.0\columnwidth]{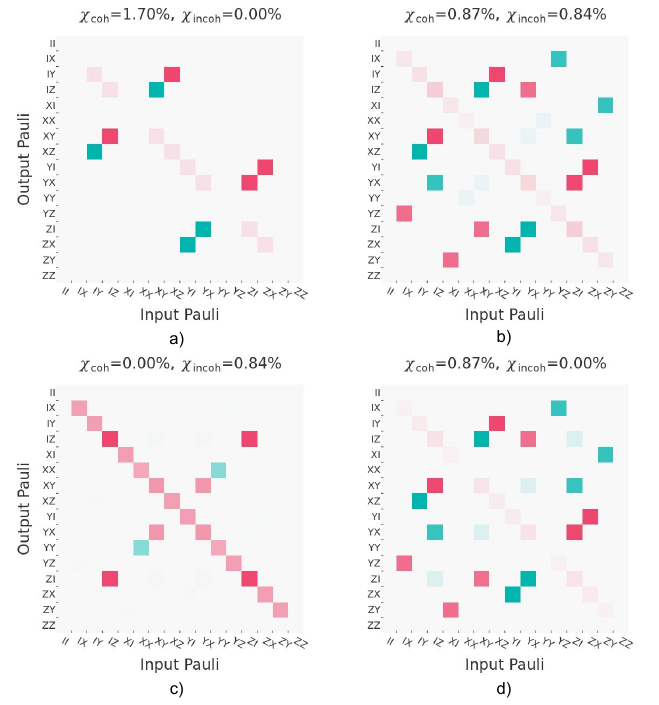}
\caption{The FSIM twirling group is applied to a RXX($\pi$/12) error. a) The Pauli transfer matrix of the initial channel is shown. The initial error channel has a coherent infidelity of $\sin^2(\pi/24) \approx 1.70\%$. b) The twirled channel has an overall infidelity of approximately $1.70\%$, with roughly equal coherent and incoherent infidelities. Note that the coherent error has also been modified by the twirling group. c) The stochastic component of the twirled channel, with a stochastic infidelity of 0.84\%. The stochastic errors include both Pauli errors on the diagonal and non-Pauli errors off the diagonal. d) The unitary component of the twirled channel, with an infidelity of 0.87\%. While the initial coherent error was $e^{-i\pi 0.042XX}$, the twirled coherent error is $e^{-i\pi (0.023XX + 0.018YY)}$.}
\label{fig:rxx-twirl}
\end{figure}

\begin{figure}[t]
\centering
\includegraphics[width=1.0\columnwidth]{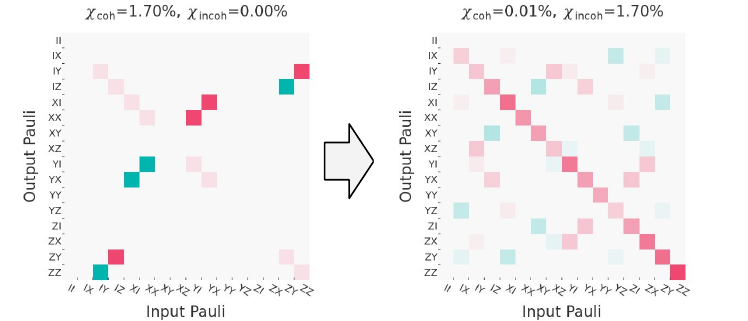}
\caption{The FSIM twirling group is applied to a RZX($\pi$/12) (CNOT-like) error. The two panels show $\Lambda-1$ for the initial and twirled error channels, where $\Lambda$ is the Pauli transfer matrix. Left: the initial error channel has a coherent infidelity of 1.70\%. Right: the twirled channel has an incoherent infidelity of 1.70\% and no coherent error. The RZX-type error is entirely tailored by the twirling group.}
\label{fig:rzx-twirl}
\end{figure}

We first examine the effect of a RZZ-type error in Fig.~\ref{fig:rzz-twirl}, and find that it is unaffected by the twirling group. We then consider a RXX-type error in Fig.~\ref{fig:rxx-twirl} and RZX-type error in Fig.~\ref{fig:rzx-twirl}, and we find that the coherent error is partially mapped and completely mapped to incoherent error, respectively.

\begin{figure}[t]
\centering
\includegraphics[width=0.8\columnwidth]{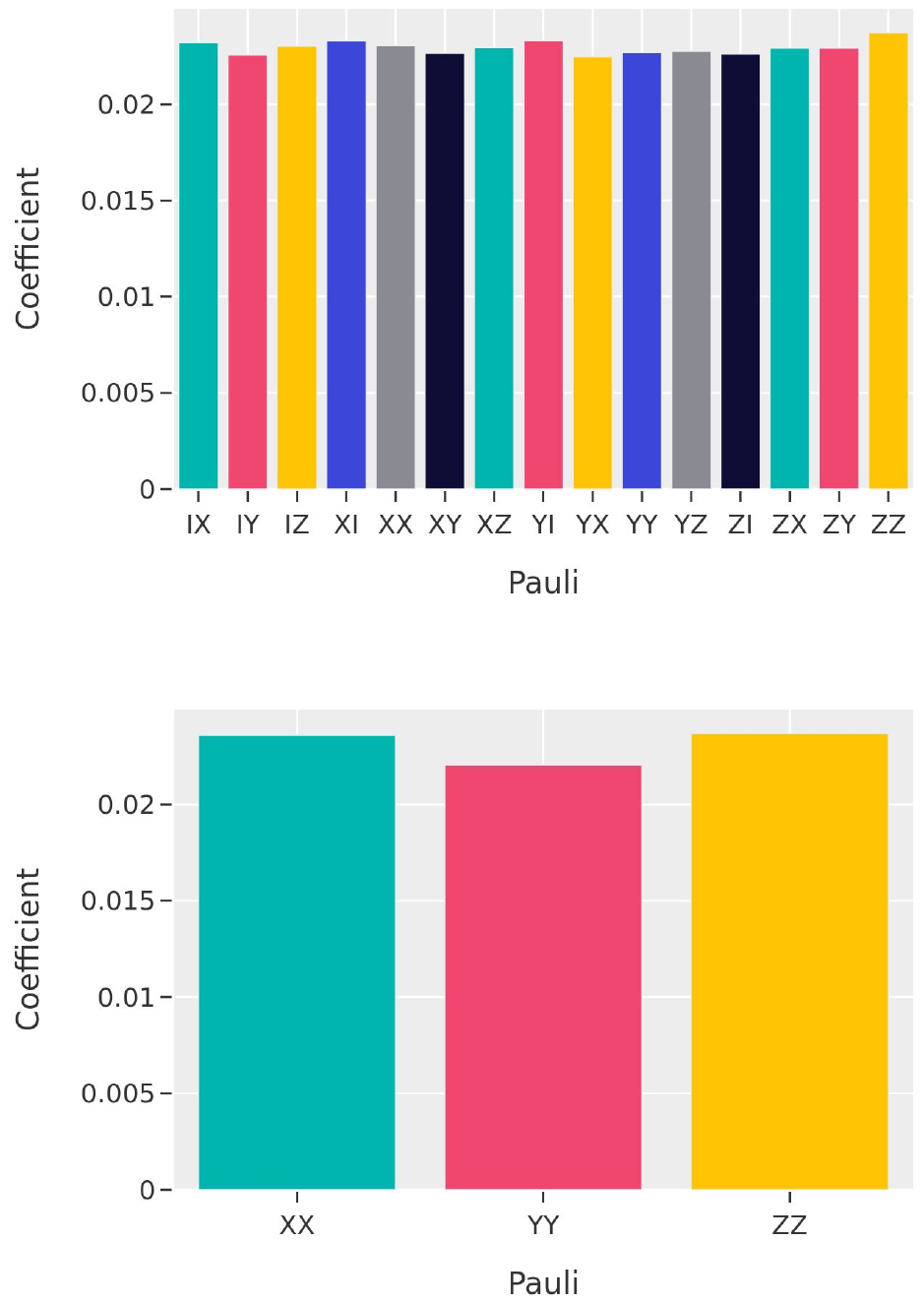}
\caption{The FSIM twirling group is applied to an ensemble of random coherent errors. Top: the average generators of the random coherent errors are plotted. Since the ensemble is Haar random, the aggregate magnitude of each generator is approximately equal, although individual instances vary randomly. Bottom: the tailored unitary generators are plotted in the same format. We observe that the remaining coherent errors have only three generators: XX, YY, and ZZ.}
\label{fig:tailored-generators}
\end{figure}

Finally, we numerically generate 1000 Haar-random coherent errors on an $\text{FSIM}(\pi/2, \pi/12)$ gate, and apply the twirling group. We find that the twirling effectiveness is approximately 80\%, meaning that the coherent error of the resulting superoperator is about 20\% of the initial value. We also observe in Fig.~\ref{fig:tailored-generators} that the remaining coherent errors are unitaries of the simplified form $e^{i(aXX +bYY + cZZ)}$.

\subsection{Clifford data regression}\label{subsec:cdr}
\begin{figure}[t]
\centering
\includegraphics[width=\columnwidth]{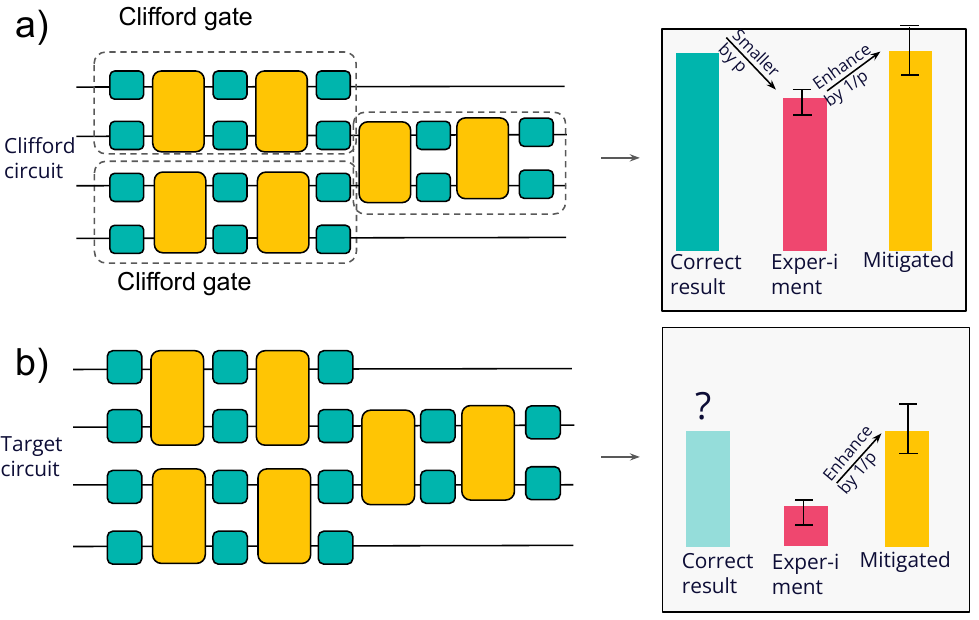}
\caption{{\bf Clifford data regression.} We execute Clifford circuits with the same structure as the target Trotter circuit, but with all the logical gates being replaced by Clifford gates. While the native gates on the hardware do not belong to the Clifford space, we express each logical Clifford gate in terms of at most two native two-qubit gates and several one-qubit gates. To mitigate errors in the target circuit that generically damp the result, we enhance the observables for the target circuit by the same amount we would in the Clifford circuit to obtain the correct result [Eq.~\eqref{eqn:cdr}].}
\label{fig:sm-cdr}
\end{figure}

Clifford data regression is a heuristic error-mitigation technique that estimates the effects of gate errors on observables as a suppression coefficient and mitigates the observables by amplifying them. The method is shown schematically in Fig.~\ref{fig:sm-cdr}

For each target circuit to be executed, we construct an ensemble of random Clifford circuits with the same native gate structure. This is so that the target circuit and the Clifford circuits are affected by noise similarly. We execute the target and Clifford circuits and measure observables $O_{\textrm{Clifford}}^{\textrm{noisy}}$ and $O_{\textrm{tgt}}^{\textrm{noisy}}$. Due to the Gottesmann-Knill theorem~\cite{gottesman1998heisenberg}, we can also exactly compute the noiseless observable in the Clifford circuits efficiently, $O_{\textrm{Clifford}}^{\textrm{noiseless}}$. Fitting $O_{\textrm{Clifford}}^{\textrm{noisy}}$ versus $O_{\textrm{Clifford}}^{\textrm{noiseless}}$, we find the average amount $r_{\textrm{suppress}}$ by which noise suppresses $\braket{O}$. The mitigated observable is then taken as
\begin{equation}
O_{\textrm{tgt}}^{\textrm{mitigated}} = O_{\textrm{tgt}}^{\textrm{noisy}}/r_{\textrm{suppress}}.
\label{eqn:cdr}
\end{equation}

Typically, the suppression factor decreases exponentially with the circuit depth; therefore, a large number of measurements are needed to accurately measure observables for deep circuits. For our error rates, the suppression factor is typically $\gtrsim 0.01$ (see Figs.~\ref{fig:sm-cdr-suppression} and~\ref{fig:sm-cdr-suppression-factors}); therefore, $\CO(10^4$) shots are sufficient.

\subsubsection{Constructing the Clifford circuits}
\begin{figure}[t]
\centering
\includegraphics[width=0.7\columnwidth]{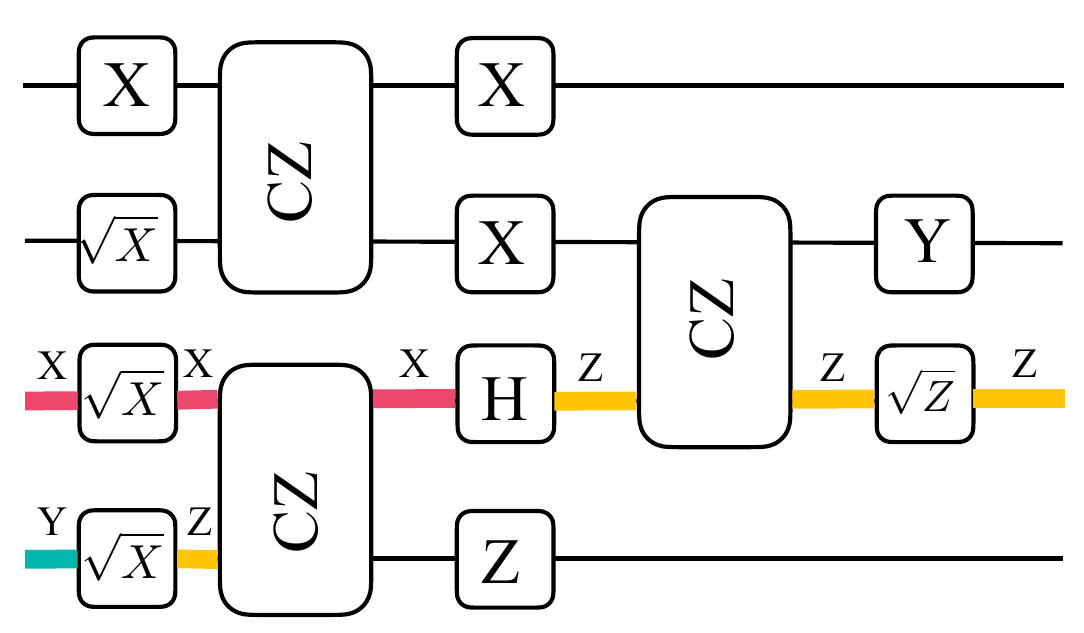}
\caption{{\bf Scheme for constructing Clifford circuits.} Given a random Clifford circuit instance and a target Pauli observable to measure, the Pauli observable is back-propagated through the circuit. The back-propagation is visualized here as a backward flow of the Pauli string, where the color scheme we use is that magenta, teal, and yellow stand for $X, Y$, and $Z$ respectively, and $I$ is not assigned any color. Each Pauli is also written on top of the flow for convenience. If the back-propagated Pauli string has any $X$ or $Y$, then that Pauli observable would be measured as 0. Here, we find $\braket{Z_3} = 0$. To make $\braket{Z_3} \neq 0$, the one-qubit gates in the first layer can be adjusted (see main text and Fig.~\ref{fig:sm-all-flows}) such that the back-propagated observable is a Pauli string with only $Z$ or $I$ on all the qubits.}
\label{fig:sm-clifford}
\end{figure}
\begin{figure}[t]
\centering
\includegraphics[width=1.0\columnwidth]{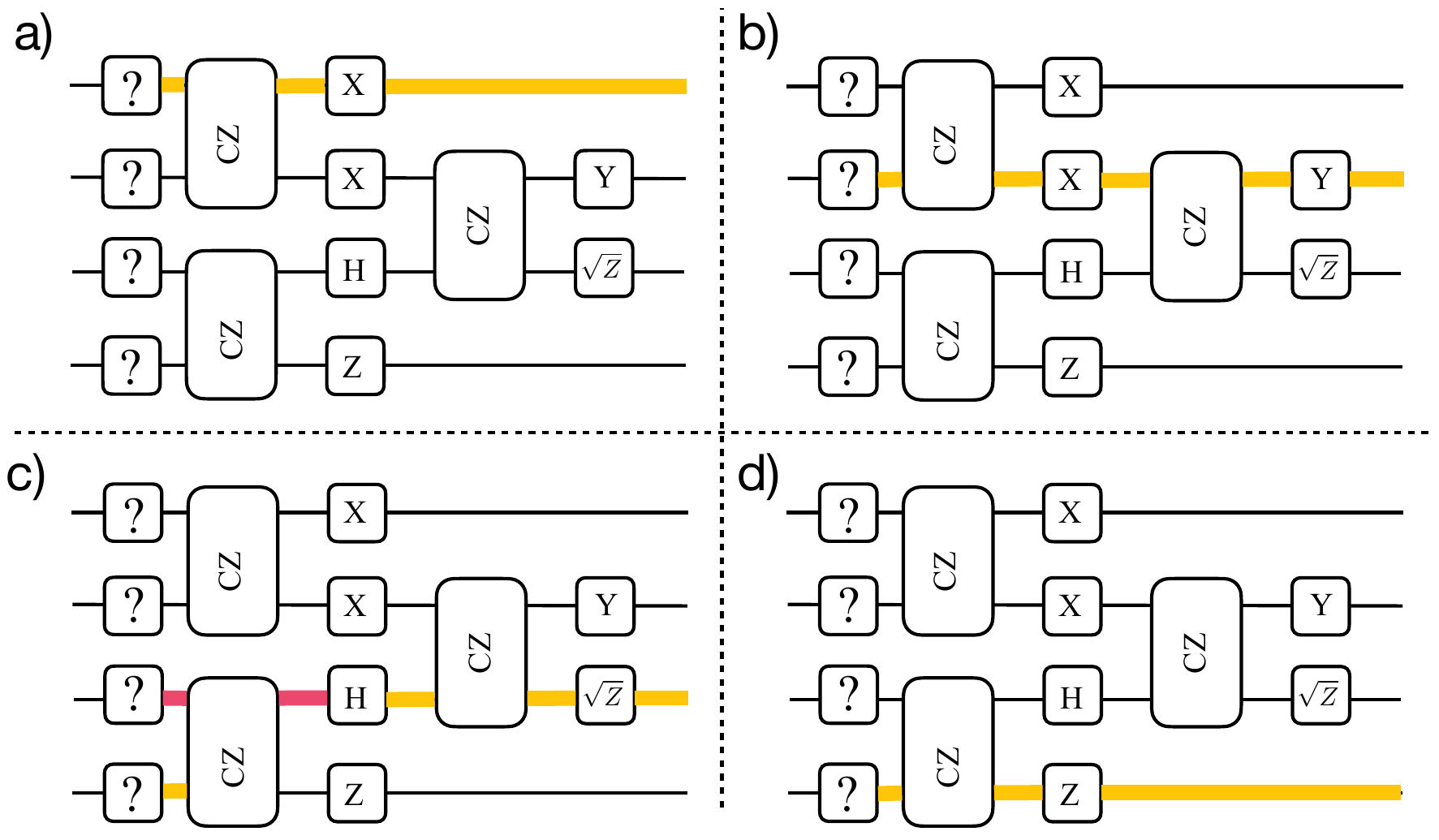}
\caption{The backward flow for the $Z$ observable on each qubit through the circuit in Fig.~\ref{fig:sm-clifford}. Since $(Z_1, Z_2, Z_3, Z_4)$ backflow to Pauli strings with no conflicts, we can pick $U_{\textrm{init}}$ which yields $\braket{Z_1}, \braket{Z_2}, \braket{Z_3}, \braket{Z_4} \neq 0$, e.g., $U_{\textrm{init}} = H_3$.} 
\label{fig:sm-all-flows}
\end{figure}

The typical expectation value of a given local Pauli observable $O$ in a random Clifford circuit is 0; however, to measure the suppression $r_{\textrm{suppress}}$ due to noise, we need a nonzero value for the noiseless expectation $\braket{O}$. Reference~\cite{czarnik2021error} describes a method to mutate Clifford circuits such that $\braket{O}\neq 0$. Here, we describe a simpler method that only needs to mutate the first layer of one-qubit gates. We assume that the first layer in our Clifford circuit consists of only one-qubit gates.

A common method for simulating Clifford circuits is to propagate the stabilizers of the initial state forward through the circuit. Here, we consider an equivalent method, which evaluates the expectation value of any Pauli observable by \textit{flowing} the Pauli observable backward through the circuit~\footnote{This flow of a Pauli through a Clifford circuit is called by various names in the literature -- as \textit{stabilizer flow} in~\cite{mcewen2023relaxing}, \textit{Pauli web} in~\cite{bombin2024unifying}, and as spacetime codes in ~\cite{delfosse2023spacetime}}. Essentially, this method simulates the evolution of the observable in the Heisenberg picture. An example is shown in Fig.~\ref{fig:sm-clifford}. We denote the Paulis $X,Y$, and $Z$ with three distinct colors -- magenta, teal, and yellow, respectively, and we ignore the sign of the Pauli although it is straightforward to track if necessary. The identity matrix is not assigned any color. 

For concreteness, let us express the random Clifford circuit as the product of the first layer of gates and all the rest of the gates, $U_{\textrm{Clifford}} = U_{\textrm{rest}} U_{\textrm{init}}$. Let $P_i$ be the local Pauli observable that we wish to measure,
\begin{equation}
\braket{P_i} = \braket{\psi_0 \vert U_{\textrm{init}}\+ U_{\textrm{rest}}\+ P_i U_{\textrm{rest}} U_{\textrm{init}} \vert \psi_0}.
\end{equation}
Without loss of generality, we assume $\ket{\psi_0} = \ket{00\cdots}$.

Here, $P_i$ is a weight-1 Pauli string at the end of the circuit. As it is back-propagated, each Clifford gate back-propagates it to another Pauli string, as exemplified in Fig.~\ref{fig:sm-clifford}. 
The expectation value $\braket{P_i}$ is nonzero \textit{iff} $P_i$ back-propagates to a Pauli string containing only $Z$ or $I$; i.e., $U_{\textrm{Clifford}}\+ P_i U_{\textrm{Clifford}}$ is a Pauli string with only $\pm Z$ or $I$. If any qubit $j$ in the back-propagated Pauli string contains any Pauli observable other than $\pm Z$ or $I$, then $\braket{P_i} = 0$; however, in this case, it is straightforward to mutate the one-qubit gate on qubit $j$ in $U_{\textrm{init}}$ such that that Pauli on that qubit becomes $\pm Z$. After adjusting the gates in $U_{\textrm{init}}$, we are left with a Clifford circuit such that $\braket{P_i} = \pm 1$.

Furthermore, rather than executing a circuit with appropriate $U_{\textrm{init}}$ for \textit{each} Pauli observable $P_i$, it would be convenient if we could find $U_{\textrm{init}}$ such that expectation values of several Pauli observables $\{ P_i\}$ are simultaneously nonzero. Thus, we are tasked with finding a layer of one-qubit gates $U_{\textrm{init}}$ such that the number of observables for which $\braket{P_i} \neq 0$ is maximized. Let us denote the Pauli string just after the first layer as $P'_i$, i.e. $U_{\textrm{rest}}\+ P_i U_{\textrm{rest}} = P'_i$. We want to find the largest group of $P'_i$ that do not have conflicting Paulis on any qubit~\footnote{Here, $X$ and $Y$ are conflicting Paulis, as are $X$ and $Z$, and $Y$ and $Z$; $X$ and $X$ are not conflicting Paulis, and so on. Note that $I$ does not conflict with any Pauli}. Once we find this group of $P'_i$, we can determine the one-qubit gates in $U_{\textrm{init}}$. To this end, we construct a graph whose nodes are $P_i$. Two nodes are connected by an edge if their $P'_i$ do not have conflicting Paulis on any qubit. Then, the largest group of $P'_i$ that do not have differing Paulis on any qubit is the largest clique on this graph. We use a clique-finding algorithm in PYTHON's NETWORKX package, and find the largest set of $P'_i$ for which this can be satisfied. We then determine $U_{\textrm{init}}$ appropriately.

We exemplify the above procedure in Fig.~\ref{fig:sm-all-flows} using the same Clifford circuit instance as Fig.~\ref{fig:sm-clifford}. The four Pauli observables $(Z_1, Z_2, Z_3, Z_4)$ get back-propagated to $(Z_1, Z_2, X_3Z_4, Z_4)$, respectively, just after $U_{\textrm{init}}$. These Pauli strings do not have any conflicts; i.e., they form a clique of size 4. Therefore we can, e.g., choose $U_{\textrm{init}} = H_3$. This will give $\braket{Z_1}, \braket{Z_2}, \braket{Z_3}, \braket{Z_4} \neq 0$.

\begin{figure}[t]
\centering
\includegraphics[width=\columnwidth]{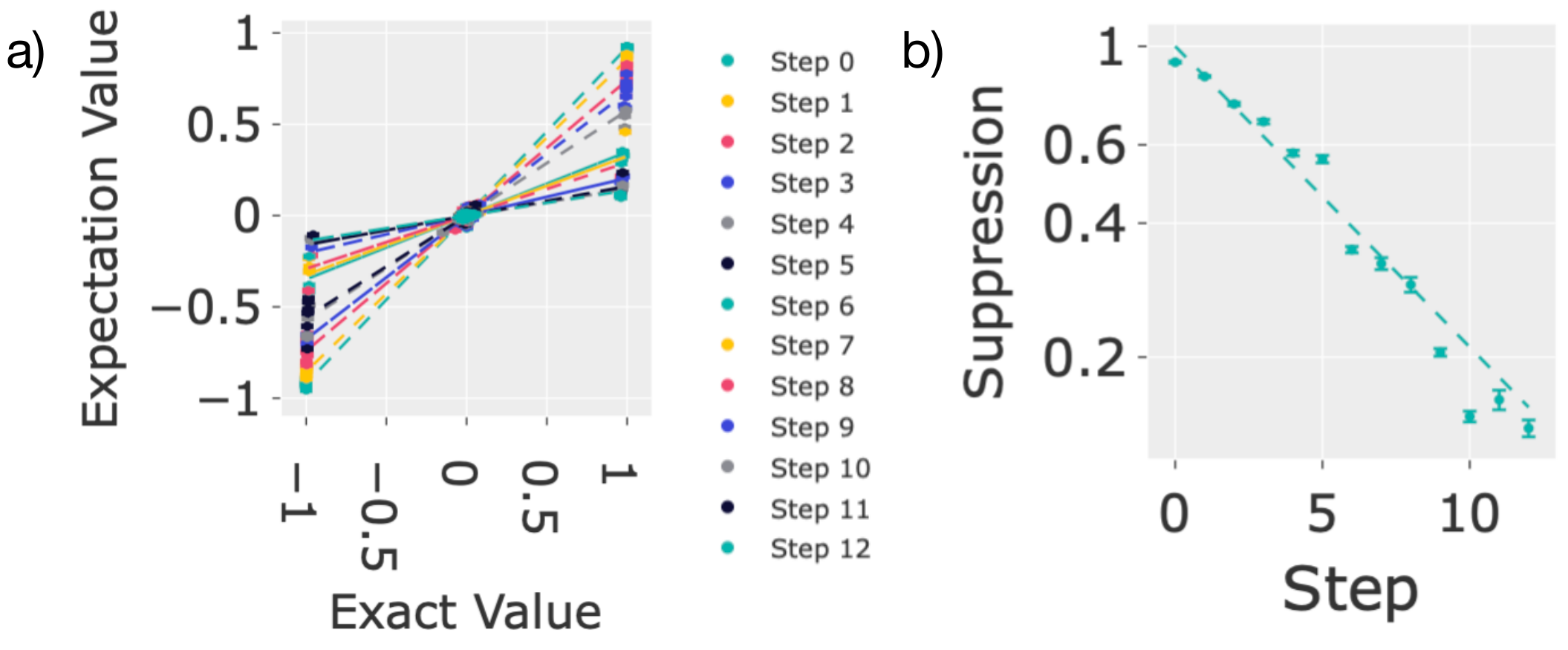}
\caption{{\bf Suppression of observables in Clifford data regression.} (a) The measured expectation values of $\braket{X}_i$ and $\braket{Y}_i$ versus their values in a noiseless experiment, for the qubit indexed as 0 in the circuit. Each color corresponds to a circuit with different depth (steps). (b) The suppression of observables in Clifford circuits, estimated from the slopes of measured versus exact expectation values in (a). This suppression factor is used to rescale the observables in the target logical circuits to mitigate their errors.}
\label{fig:sm-cdr-suppression}
\end{figure}

\begin{figure}[t]
\centering
\includegraphics[width=\columnwidth]{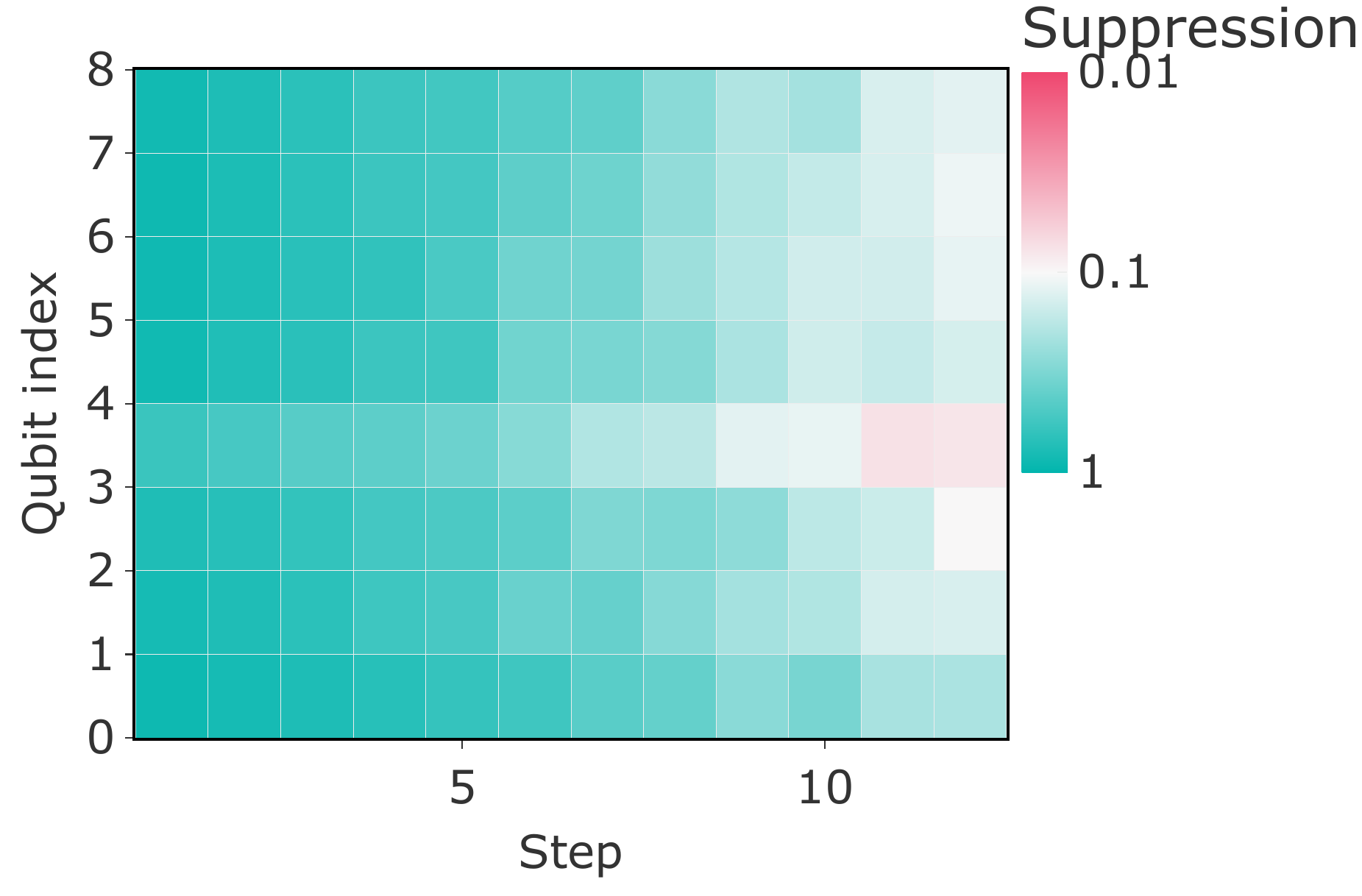}
\caption{{\bf Suppression of observables in Clifford data regression.} The suppression of observables in Clifford circuits versus qubit index and circuit depth. These suppression factors are used to rescale the observables in the target logical circuits, to mitigate the error in the target logical circuits. Panel (b) in Fig.~\ref{fig:sm-cdr-suppression} shows the trace for qubit 0. }
\label{fig:sm-cdr-suppression-factors}
\end{figure}

\subsection{Conservation of total spin}\label{subsec:Zconservation}
The spin Hamiltonian $H$ [Eq.~\eqref{eq:H}] conserves $\sigma^z_{\textrm{tot}} = \sum_i \sigma^z_i$. Therefore, in the experiments with a propagating wave packet [Fig.~\ref{fig3}], where the initial state is an eigenstate of $\sigma^z_{\textrm{tot}}$ with eigenvalue $N-2$, the quantum state after any number of Trotter steps still remains an eigenstate of $\sigma^z_{\textrm{tot}}$ with eigenvalue $N-2$.

In practice, due to hardware noise, we also measure bitstrings with $\sigma^z_{\textrm{tot}} \neq N-2$. Postselecting for bitstrings with $\sigma^z_{\textrm{tot}} = N-2$ is not necessarily scalable. We instead enforce conservation of $\sigma^z_{\rm tot}$ on an \textit{ensemble} level by rescaling each expectation value $\braket{\sigma^z_i}$ with $(N-2)/\braket{\sigma^z_{\rm tot}}$.

\section{Additional experimental data}
\begin{figure*}[t]
\centering
\includegraphics[width=1.0\textwidth]{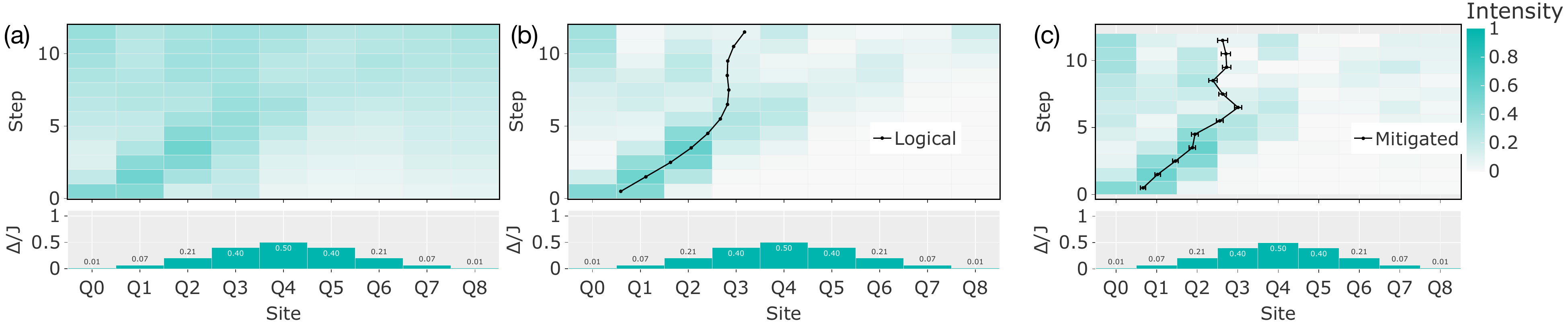}
\caption{Additional data from the experiment with an electromagnetic wave incident on an inhomogeneous plasma with $\Egap_{\textrm{max}}=\Ehop/2$ from a region with $\Egap=0$. Wave-packet densities (a) measured in the experiment without error mitigation, (b) after error mitigation on the experimental data, and (c) in a noiseless simulation of the circuits.}
\label{fig:sm-mass-barrier}
\end{figure*}

Figures~\ref{fig:sm-mass-barrier}(a) and ~\ref{fig:sm-mass-barrier}(b) show the raw and error-mitigated experimental data, respectively, for a wave packet incident on a plasma with $\Egap_{\textrm{max}}/\Ehop=1/2$, and the results of the noiseless simulation are given in Fig.~\ref{fig:sm-mass-barrier}(c). The plasma barrier ${\rm max}\ \Egap=\Ehop/2$ is larger than the incident wave packet's average energy $\epsilon_k=0$; therefore, it is an overdense plasma. In our experiment, the $\ket{0}$ state of the qubit corresponds to wave-packet density 1 and the $\ket{1}$ state corresponds to wave packet density 0; an equal incoherent mixture (or coherent superposition) corresponds to density 1/2. One of the dominant error sources in the experiment is the qubits' natural tendency to decay to the $\ket{0}$ state on a time scale $T_1 \sim 37.6\mu$s (see Table~\ref{tab:ankaa-3-char}), which is roughly 50$\%$ longer than the duration of the deepest circuits. This effect, which manifests as an amplitude-damping noise channel, gets partially twirled in our circuits. It is reasonable to expect that the twirling drives the qubits to approximately an incoherent mixture of $\ket{0}$ and $\ket{1}$, which has a uniform wave-packet density of 1/2, as can be observed in Fig.\ref{fig:sm-mass-barrier}(a). Removing this bias is crucial to obtaining accurate results, and it is qualitatively accomplished by Clifford data regression, as evidenced by the qualitatively better agreement between Figs.\ref{fig:sm-mass-barrier}(b) and (c).

\begin{figure*}[t]
\centering
\includegraphics[width=1.0\textwidth]{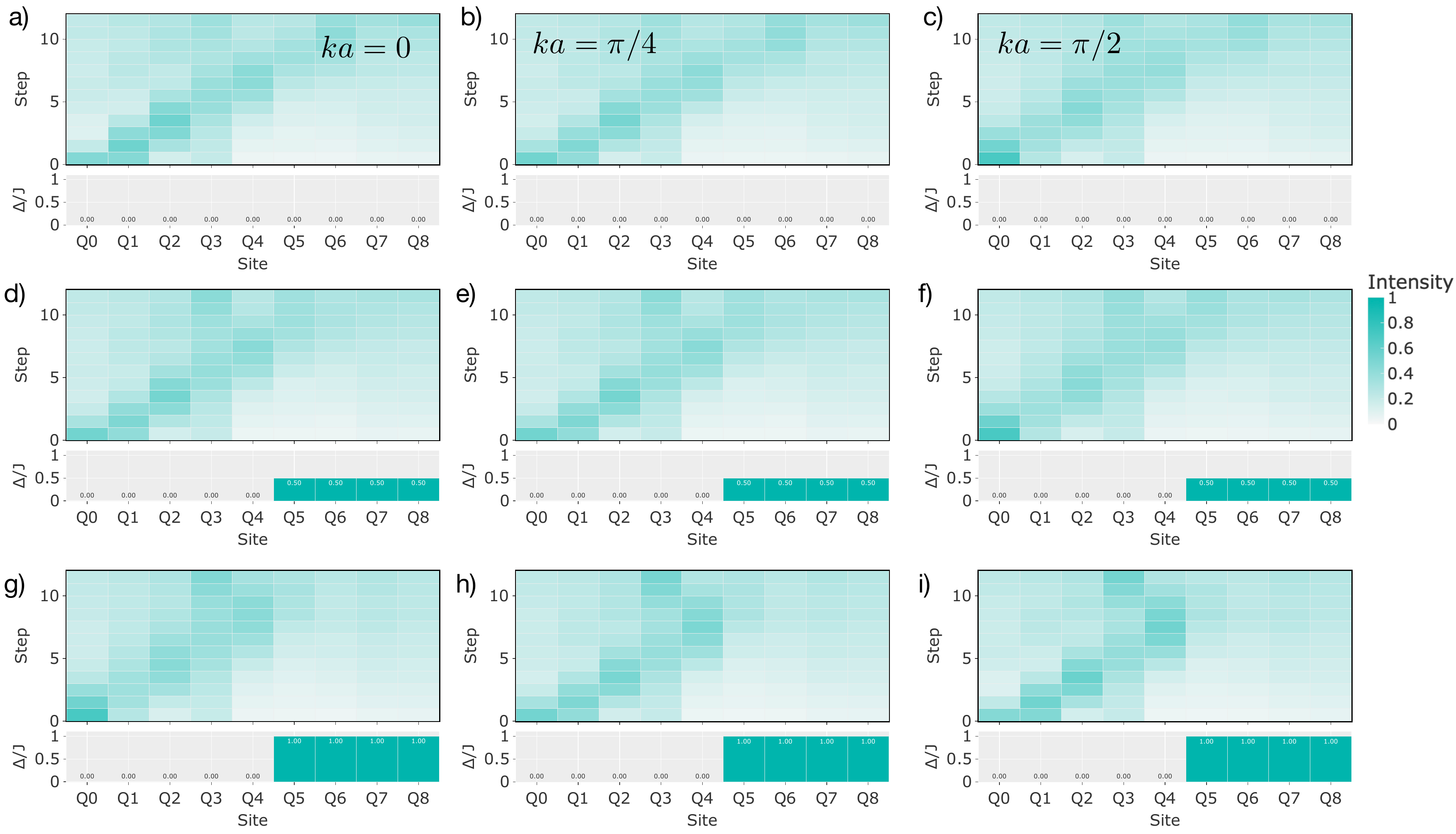}
\caption{Additional raw data from experiments with an electromagnetic wave incident on a plasma with $\Egap\neq 0$ from a region with $\Egap=0$, without error mitigation. In the first column, the initial wave packet is prepared with a relative phase between the two sites as $ka=0$; this phase is $ka=\pi/4$ in the second column and $\pi/2$ in the third column. The wave packet is incident on a plasma with $\Egap/\Ehop = 0$ in the first row, $\Egap/\Ehop = 0.5$ in the second row, and $\Egap/\Ehop = 1$ in the third row.}
\label{fig:sm-reflection}
\end{figure*}

\begin{figure*}[t]
\centering
\includegraphics[width=1.0\textwidth]{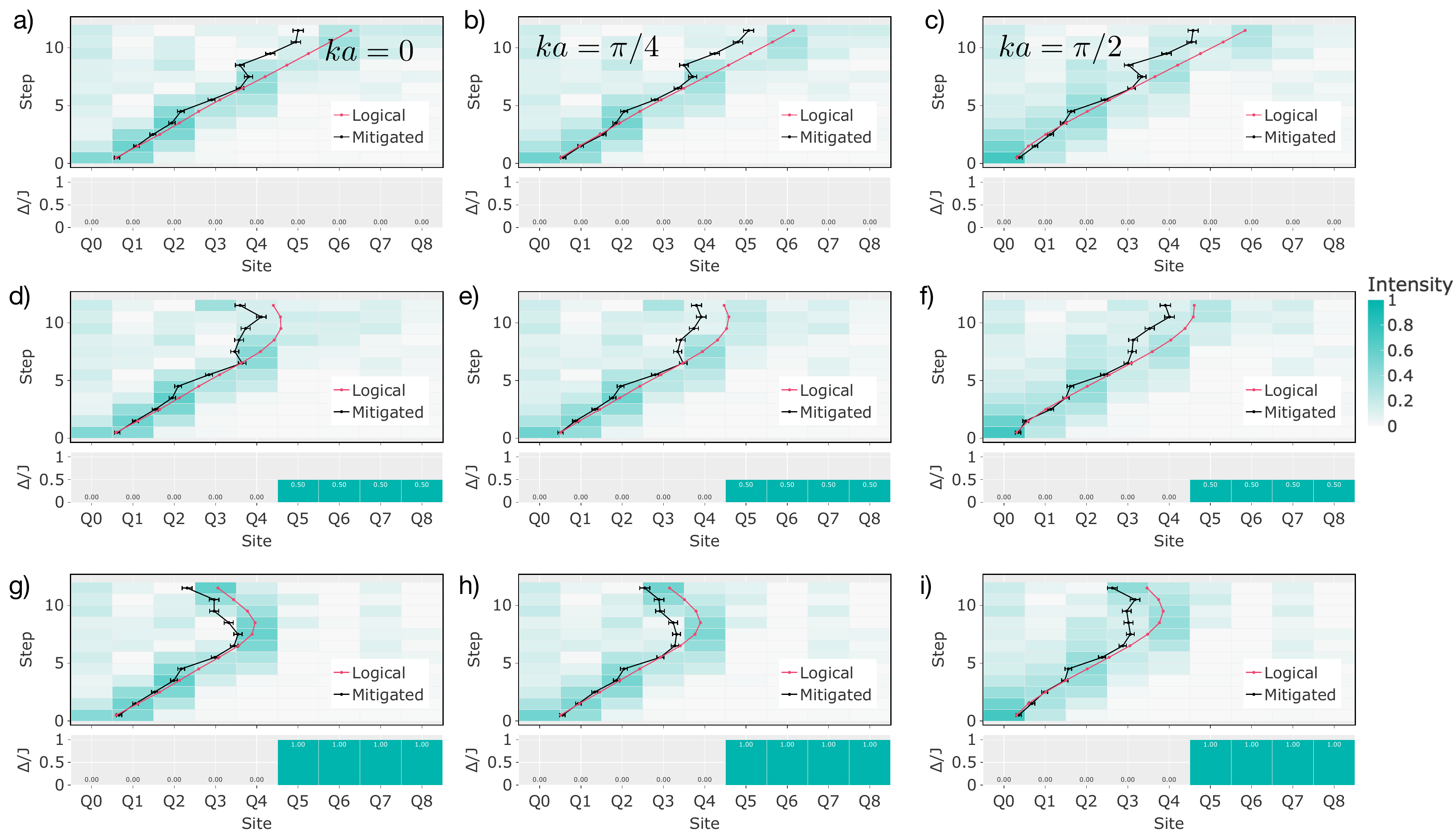}
\caption{Wave-packet densities in the experiments of Fig.~\ref{fig:sm-reflection} after implementing error mitigation on the data. The order of the panels is the same as in Fig.~\ref{fig:sm-reflection}.}
\label{fig:sm-reflection-mitigated}
\end{figure*}

Figure~\ref{fig:sm-reflection} shows the raw experimental data for wave-packet propagation in vacuum ($\Egap/\Ehop=0$) or incident on a plasma ($\Egap/\Ehop \neq 0$), starting with different relative phases between the wave-packet amplitudes. In principle, the relative phase sets the wave vector $k$, which determines the packet's group velocity. In practice, this fact is not so visible in our small system, which can only initiate a relatively narrow wave packet that is not monochromatic. Figure~\ref{fig:sm-reflection-mitigated} shows that the results are significantly improved after implementing Clifford data regression. In Figs.~\ref{fig:sm-reflection} and~\ref{fig:sm-reflection-mitigated}, the plasma is underdense in panels~(e) and (f), where the barrier height $\Egap=\Ehop/2$ is smaller than the wave packet's average energy of $\epsilon_k = \Ehop/\sqrt{2}$ and $\epsilon_k = \Ehop$, respectively. The plasma's density is critical in panel (i), with $\epsilon_k = \Egap = \Ehop$.

Figures~\ref{fig:large} and~\ref{fig:reflection_coeff} show that a larger system can reliably initiate nearly monochromatic wave packets and probe the $k$ dependence of the reflection of the wave packets at the plasma boundary.

\section{Reflection at a sharp jump in the plasma density}\label{app:reflection_coefficient}
Reflection and transmission of EM waves at a sharp boundary is the simplest case in which the plasma density is nonuniform.

Reflection and transmission coefficients are well defined in a thermodynamically large system where a plane wave is incident on a sharp boundary, with some of it reflected and some transmitted. Analogous to the eigenmodes with hard walls in Appendix~\ref{app:exact_soln}, the plane-wave eigenmodes in a uniform thermodynamically large system have amplitudes
\begin{align}
c_{jk} = \sqrt{\frac{\Egap+\hbar\omega_k}{8\hbar\omega_k}}\exp(i kja), \quad\ \textrm{if}\ j\ \textrm{is odd} \nonumber\\
c_{jk} = \frac{\sin ka}{\sqrt{8\hbar\omega_k(\Egap+\hbar\omega_k)}}\exp(i kja), \quad\ \textrm{if}\ j\ \textrm{is even}.
\end{align}

We suppose an incoming wave with amplitude 1 hits the sharp density jump at $j=0$ from the left, i.e. from $j<0$. 
Here, we assume an EM wave is incident from the left of the boundary. The boundary is at $j=0$, and the plasma density is such that $\Egap=0$ when $j<0$. 
We denote the reflected wave as having amplitude $r$ and the transmitted wave as having amplitude $t$. Thus, the spatial amplitudes of the plane wave to the left ($j<0$) are
 \begin{align}
c_{jk_{\textrm{in}}} = \frac{1}{\sqrt{8}} \left(\exp(i k_{\textrm{in}}ja) + r \exp(-i k_{\textrm{in}}ja)\right), \ \textrm{if}\ j\ \textrm{is odd}, \nonumber\\
c_{jk_{\textrm{in}}} = \frac{\sin {k_\textrm{in}a}}{\sqrt{8}\hbar\omega_{k_\textrm{in}}} \left(\exp(i k_{\textrm{in}}ja) - r \exp(-i k_{\textrm{in}}ja)\right), \ \textrm{if}\ j\ \textrm{is even}
\label{eqn:incident_and_reflected_waves}
\end{align}
where $\hbar\omega_{k_\textrm{in}} = |\sin {k_\textrm{in}a}|$, and that of the transmitted wave ($j\geq0$) is
\begin{align}
c_{jk_{\textrm{out}}} = t\sqrt{\frac{\Egap+\hbar\omega_{k_{\textrm{out}}}}{8\hbar\omega_{k_{\textrm{out}}}}} \exp(i k_{\textrm{out}}ja), \quad\ \textrm{if}\ j\ \textrm{is odd} \nonumber\\
c_{jk_{\textrm{out}}} = t \frac{\sin ka}{\sqrt{8}\hbar\omega_{k_\textrm{out}}} \exp(i k_{\textrm{out}}ja), \quad\ \textrm{if}\ j\ \textrm{is even}
\label{eqn:transmitted_wave}
\end{align}
where $k_{\textrm{out}}$ is related to $k_{\textrm{in}}$ by energy conservation,
\begin{equation}
\hbar\omega_{k_\textrm{out}} \equiv \sqrt{\Ehop^2\sin^2k_\textrm{out}a+\Egap^2} = \hbar\omega_{k_{\textrm{in}}} = \Ehop\sin k_{\textrm{in}}a.
\end{equation}
Hereafter we denote $\hbar\omega \equiv \hbar\omega_{k_\textrm{out}} = \hbar\omega_{k_\textrm{in}}$. We denote $c_{jk_{\textrm{in}}}$ and $c_{jk_{\textrm{out}}}$ as simply $c_j$.

At the boundary, we have the relations:
\begin{align}
& \hbar\omega c_0 = \frac{i\Ehop}{2}(c_1 - c_{-1}) - \Egap c_0 \nonumber\\
& \hbar\omega c_{-1} = \frac{i\Ehop}{2}(c_0 - c_{-2}).
\end{align}

Solving for $r$ using the expressions for $c_{j<0}$ and $c_{j\geq0}$ in Eqs.~\eqref{eqn:incident_and_reflected_waves} and~\eqref{eqn:transmitted_wave}, we obtain
\begin{align}
r = \frac{ (e^{ik_{\textrm{in}}a}-2i\frac{\Egap}{\Ehop})\sqrt{\sin k_{\textrm{in}}a+\frac{\Egap}{\Ehop}}-e^{ik_{\textrm{out}}a}\sqrt{\sin k_{\textrm{in}}a-\frac{\Egap}{\Ehop}} }{ (e^{-ik_{\textrm{in}}a}+2i\frac{\Egap}{\Ehop})\sqrt{\sin k_{\textrm{in}}a+\frac{\Egap}{\Ehop}}+e^{ik_{\textrm{in}}a}\sqrt{\sin k_{\textrm{in}}a-\frac{\Egap}{\Ehop}} }.
\end{align}

\begin{figure}[t]
\centering
\includegraphics[width=0.7\columnwidth]{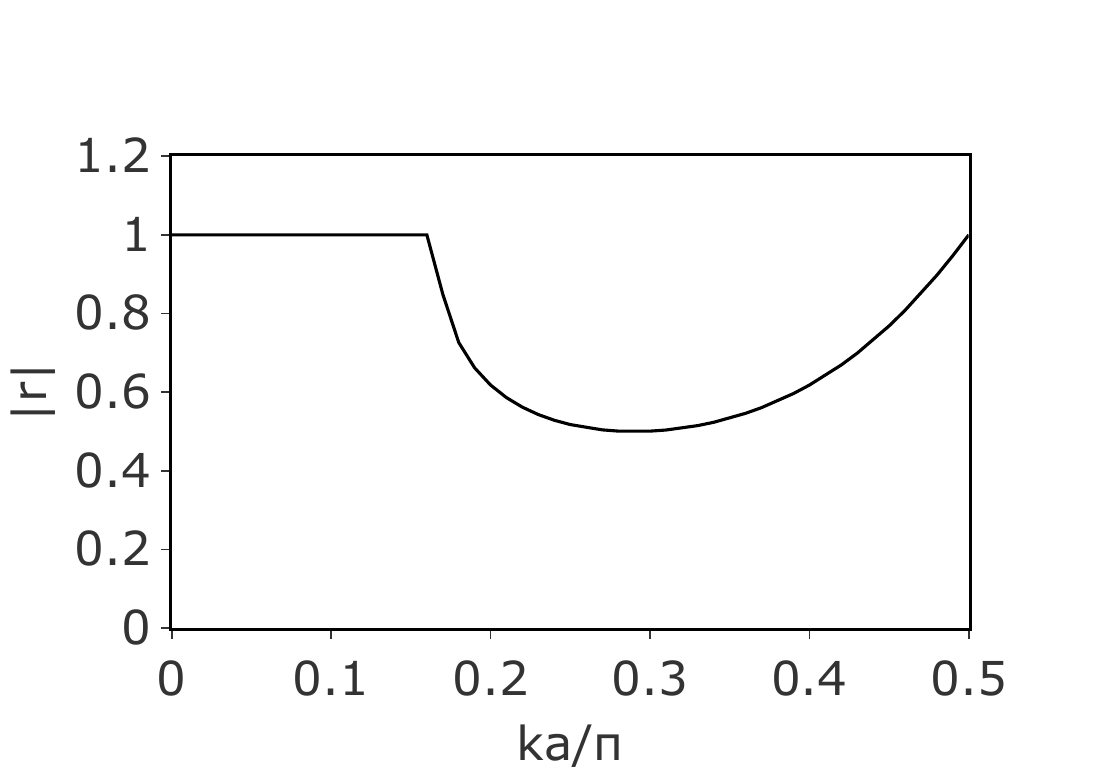}
\caption{Reflection coefficient vs $k$ at $\Egap=\Ehop/2$.}
\label{fig:reflection_coeff}
\end{figure}

The reflection coefficient $|r|$ versus $k_{\textrm{in}}$ is plotted in Fig.~\ref{fig:reflection_coeff}. When $|\sin ka|<\frac{\Egap}{\Ehop}$, $r$ is a pure phase; therefore, the incident wave is entirely reflected, $|r| = 1$. When $|\sin ka|>\frac{\Egap}{\Ehop}$, the incident wave is partially transmitted, i.e. $|r| < 1$.

\bibliography{bibliography}
\end{document}